\documentclass[amsmath,amssymb,floatfix,prb,twocolumn]{revtex4}

\usepackage{enumerate}
\usepackage{graphicx}
\usepackage{txfonts}
\usepackage{natbib}

\newcommand{\boldnabla}{\mbox{\boldmath$\nabla$}}

\begin{document}

\title{Consequences of spontaneous reconnection at a two-dimensional non-force-free current layer}
\author{J. Fuentes-Fern\'andez}
\email{jorge@mcs.st-andrews.ac.uk}
\author{C.~E. Parnell}
\author{A.~W. Hood}
\author{E.~R. Priest}
\affiliation{Department of Mathematics and Statistics, University of St Andrews, North Haugh, St Andrews, KY16 9SS, United Kingdom} 
\author{D.~W. Longcope}
\affiliation{Department of Physics, Montana State University, Bozeman, Montana 59717, USA} 

\begin{abstract}
Magnetic neutral points, where the magnitude of the magnetic field vanishes locally, are potential locations for energy conversion in the solar corona. The fact that the magnetic field is identically zero at these points suggests that for the study of current sheet formation and of any subsequent resistive dissipation phase, a finite beta plasma should be considered, rather than neglecting the plasma pressure as has often been the case in the past. The rapid dissipation of a finite current layer in non-force-free equilibrium is investigated numerically, after the sudden onset of an anomalous resistivity. The aim of this study is to determine how the energy is redistributed during the initial diffusion phase, and what is the nature of the outward transmission of information and energy. The resistivity rapidly diffuses the current at the null point. The presence of a plasma pressure allows the vast majority of the free energy to be transferred into internal energy. Most of the converted energy is used in direct heating of the surrounding plasma, and only about 3\% is converted into kinetic energy, causing a perturbation in the magnetic field and the plasma which propagates away from the null at the local fast magnetoacoustic speed. The propagating pulses show a complex structure due to the highly non-uniform initial state. It is shown that this perturbation carries no net current as it propagates away from the null. The fact that, under the assumptions taken in this paper, most of the magnetic energy released in the reconnection converts internal energy of the plasma, may be highly important for the chromospheric and coronal heating problem.
\end{abstract}

\maketitle


\section{Introduction}

Magnetic reconnection is likely to play an important role in the coronal heating problem, but the relative importance of reconnection and wave heating models in the different regions of the corona is still unclear \cite{Walsh03,Klimchuk06,Hood10}. Magnetic null points are important locations for magnetic reconnection. Their relevance in coronal studies has been stated by numerous authors \cite{Ugarte07,Longcope09,Masson09}, and, in two-dimensions, they are the only locations where reconnection can occur. Two big questions about magnetic reconnection at null points are: into what forms of energy is the stored magnetic energy converted, and where does this energy conversion take place. It is well known that magnetic energy is transferred into internal, kinetic and fast particle energy, but the question of the proportions in each is still unsettled.

At two-dimensional magnetic X-points in particular, the answers to all these questions, as well as the source of the required magnetic energy for reconnection, have been a subject of study for decades. In many situations in the solar corona, the continuous slow photospheric motions of the magnetic footpoints feed energy to the magnetic field. This energy may be stored in the form of many current density layers \cite{Parker72,vanBallegooijen85}. Locally within these current layers, the properties of the plasma may at some point reach some appropriate conditions for which reconnection becomes important. For instance, current layers may undergo current induced microinstabilities creating an anomalous resistivity \cite{Galeev84,Raadu88,Yamada97}, which then permits the dissipation of the current via reconnection. In this process, part of the magnetic energy is transferred into kinetic and/or internal energy of the plasma (plus particle acceleration in kinematic models). This is likely to be the case in many examples of solar flares \cite{Antiochos82,Barta11}, and is a possible mechanism for coronal heating via myriads of small-scale nano-flares \cite{Galsgaard96}. The onset of these instabilities, as well as the energy partitioning, are still not well understood.

In this paper, we consider the case where a current layer has been slowly formed in a non-force-free scenario, using a magnetohydrodynamic (MHD) model. We apply a local enhancement in resistivity in the current layer to mimic the sudden onset of an anomalous resistivity causing spontaneous magnetic reconnection (without the presence of any external flow \cite{Longcope94,Craig94}). This approach fixes the amount of available magnetic energy in the process, and is completely different to steady state reconnection models such as the fast reconnection regimes \cite{Petschek64,Biskamp86,Priest86}, where the rate with which the reconnection is being driven is the same as the actual rate with which the flux is reconnected.

A recent study which considered the source and nature of the energy conversion associated with 2D spontaneous reconnection has been carried out in \citet{Longcope07} (from now on referred to in the paper as {\bf LP07}). In this paper, they make an analytical study of the fast magnetosonic (FMS) wave launched by reconnection in a current sheet after a sudden increase in the resistivity. By conducting a one-dimensional analysis in which they investigate the leading order term only, which they assume to be $m=0$, they found that a propagating sheath of current which travels out from the null at the local Alfv\'en speed, converting the magnetic energy into kinetic energy as it moves through the volume. They start from an ideal Green's current sheet \cite{Green65}, where the field is potential everywhere except in an infinitesimally thin, but finite length, current sheet with an infinite current density. The dissipation of the current is then modelled by introducing a uniform diffusivity, $\eta$, everywhere in the domain, which is big enough for their dynamics to be approximated by the linear resistive MHD equations. An important assumption is that they neglect the plasma pressure, and hence, treat the problem as purely magnetic.

Their analytical solution has two distinct parts: the first is described by a diffusion equation and the second by a propagating wave equation. As soon as the resistivity is enhanced, diffusion expands the current rapidly (at a rate much quicker than any wave travel time), and the current expansion then slows down to the point where the speed of expansion couples to the local FMS mode. From that point on, a propagating fast wave expands outwards from the edge of the diffusion region, carrying most of the energy converted in the diffusion process.

The coupling between diffusion and FMS waves after a sudden enhanced resistivity was first studied \cite{Forbes82} for an infinite current sheet embedded in a uniform external field. The work of LP07 is a generalization from the uniform magnetic field to a more complex X-point scenario.

In their derivations, several assumptions are made, which we summarise here:\\
1) their analytical study is conducted within a linear regime, where the natural dissipation length scale ($l_{\eta}\sim \eta^{1/2}$) is much larger than the size of their current sheet;\\
2) they neglect the effects of plasma pressure, for simplicity;\\
3) they assume that the most significant changes in the magnetic field are in the $m=0$ term. Hence, the perturbation has cylindrical symmetry, with a circular expanding sheath.

The overall results they find are:\\
1) after the initial diffusive phase, the magnetic energy is converted almost entirely into kinetic energy;\\
2) their outward propagating wave carries a net positive current, due to the diminished current left behind by the reconnection;\\
3) a persistent peak of current density remains at the location of the X-point, where a small amount of energy dissipation continues, driven by a velocity inflow set off after the outgoing FMS wave has past. Thus, within the diffusion region around the X-point, a steady flow consistent with reconnection is left behind: outwards parallel to the axis of the current sheet and inwards perpendicular to it;\\
4) in addition to the outwards propagation of the wave, to the right and left of the null the flow field encroaches slowly inwards as the pulse expands;\\
5) the $m=0$ component of the FMS wave is a concentric sheath of current that propagates outwards and is narrow but not infinitesimal. Its width is proportional to the radius, and it carries a net current with it, so that, once the FMS wave has passed by, the field inside the sheath has less current and is much closer to potential.

In the present paper, we conduct numerical experiments to consider the same questions as in the analytical treatment of LP07. However, we cannot start our numerical experiments from the same initial condition. The main reason for not using a Green-type initial current sheet in our numerical experiments is that, with a finite grid (of any resolution) the width of the current sheet becomes finite, and the current layer is no longer in equilibrium. Also, inside the current sheet, where the magnetic field reverses sign, the magnetic pressure varies across the current sheet, creating an unbalanced magnetic pressure force. To analyse the problem of spontaneous reconnection and the nature of any waves launched, it is essential that our experiments start from a genuine non-force-free magnetohydrostatic (MHS) equilibrium, where the Lorentz forces of the central current layer are entirely balanced by plasma pressure gradients.

We use full 2D MHD numerical simulations, starting with an initial equilibrium X-point that contains a main elongated current layer at the location of the null, which has a high, but not infinite peak. The current also extends faintly along the four separatrices of the system, such that the current structures are able to contain non-zero plasma pressure gradients. This initial state is given by \citet{Fuentes11} (which will be referred to as {\bf FF11}), as a numerical solution to the non-resistive collapse of a squashed potential X-point, achieved by viscous relaxation. This state is a ``quasi-static'' equilibrium in which force balance is achieved throughout the domain, except locally at the null and along the separatrices. At these locations, a very slow asymptotic regime towards an infinite time singularity remains and is numerically and physically unavoidable.

For further details of work done regarding current sheet formation around 2D X-points for both force-free and non-force-free solutions, see FF11. In the first paper of that series \cite{Fuentes10}, the authors studied the role of plasma pressure effects in simple hydromagnetic scenarios. Starting from the non-force-free ``quasi-equilibrium'' from FF11, we study the impulsive reconnection due to an anomalous localised resistivity in a narrow diffusion region at the location of the null. We aim to compare our numerical results with the simplified analytical study of LP07.

The paper is structured as follows. In Sec. \ref{sec:sec1}, we define the equations governing the dynamical evolution. In Sec. \ref{sec:sec2}, we present the initial equilibrium state to start our numerical experiment, mentioning its properties, its relevance, and its origin. The main results are shown in Sec. \ref{sec:sec3}, followed by a general discussion in Sec. \ref{sec:sec4}.


\section{Resistive MHD equations} \label{sec:sec1}

\begin{figure*}[t]
  \centering
  \includegraphics[scale=0.31]{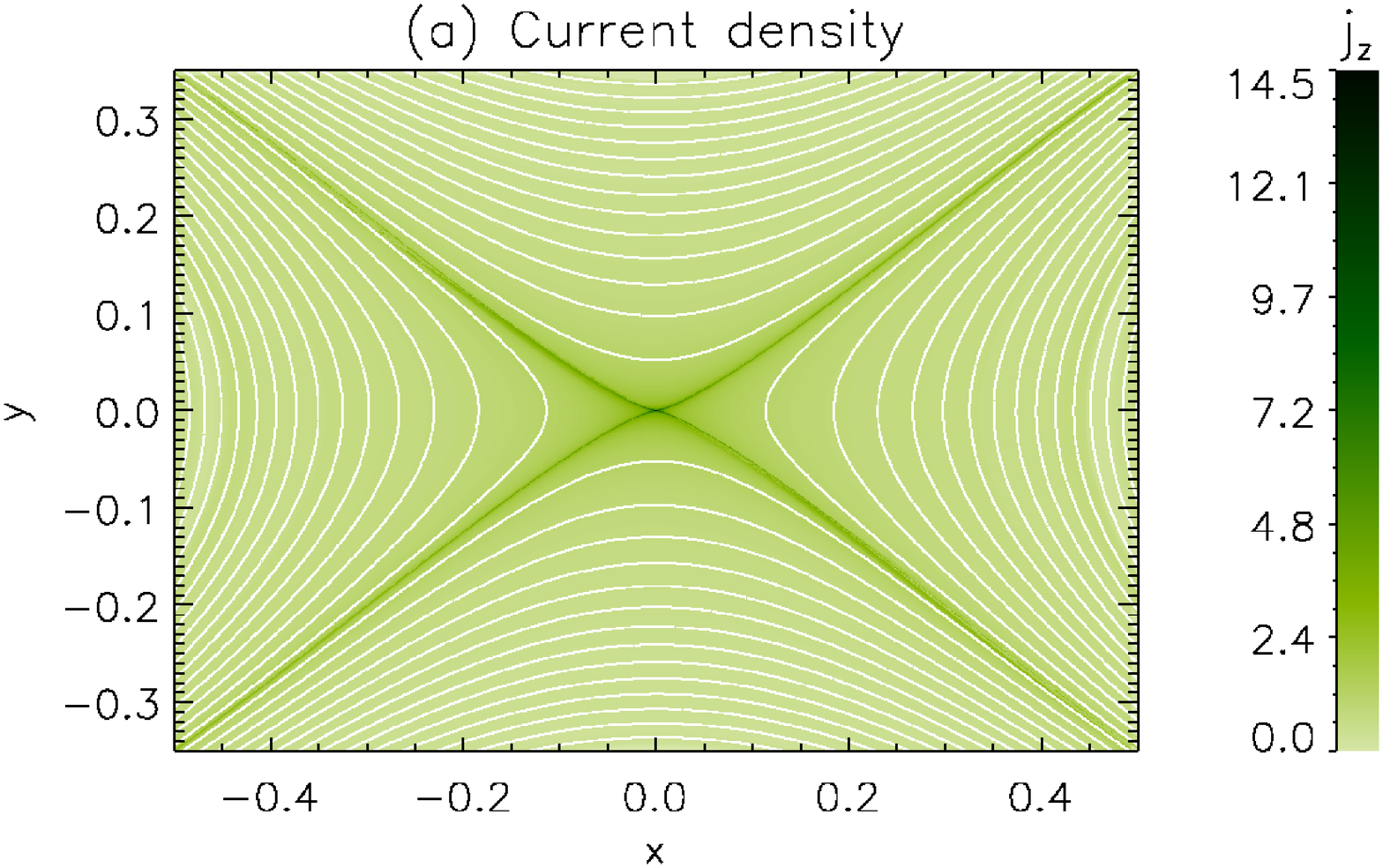}
  \includegraphics[scale=0.31]{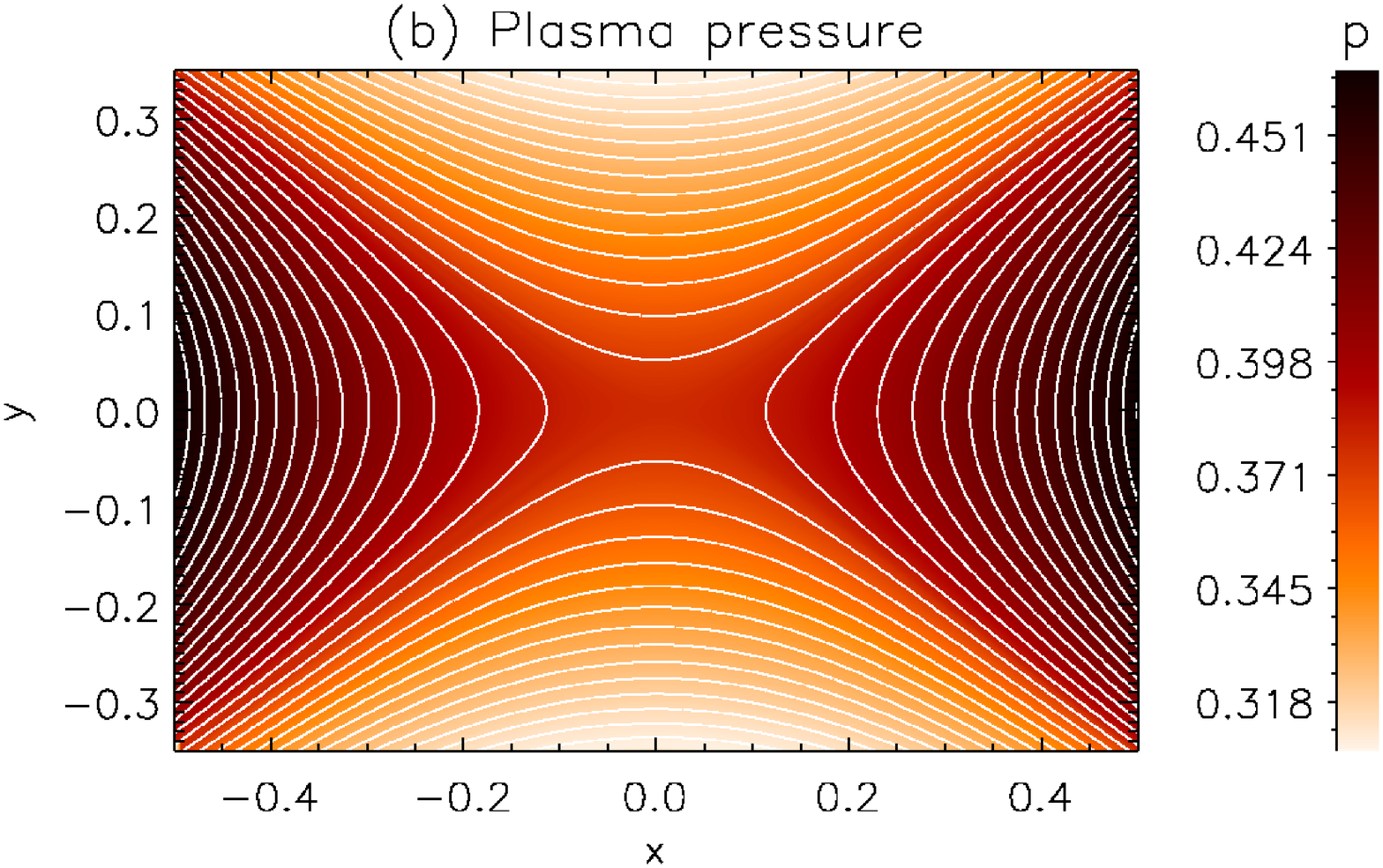}
  \caption{Two-dimensional contour plots of (a) electric current density and (b) plasma pressure for the initial equilibrium state. Solid white lines are the magnetic field lines, drawn as contours of the flux function, $A_z$, where $A_z<0$ outside the cusp (top and bottom), $A_z>0$ inside the cusp (left and right) and $A_z=0$ at the null and on the separatrices. This state is taken from the final equilibrium obtained after the dynamical non-resistive relaxation of FF11.}
  \label{fig:initial}
\end{figure*}

For the numerical experiments studied in this paper, we have used Lare2D \cite{Arber01}, a staggered Lagrangian-remap code with user controlled viscosity and resistivity, that solves the full MHD equations. The numerical code uses the normalised MHD equations, where the normalised magnetic field, density and lengths,
\begin{eqnarray*}
x=L\hat{x}\;,\;\;\;y=L\hat{y}\;,\;\;\;{\bf B}=B_n\hat{\bf B}\;,\;\;\;\rho=\rho_n\hat{\rho}\;,
\end{eqnarray*}
imply that the normalising constants for pressure, internal energy, current density and plasma velocity are,
\begin{eqnarray*}
p_n=\frac{B_n^2}{\mu}\;,\;\;\;\epsilon_n=\frac{B_n^2}{\mu\rho_n}\;,\;\;\;j_n=\frac{B_n}{\mu L}\;\;\;{\rm and}\;\;\;v_n=\frac{B_n}{\sqrt{\mu \rho_n}}\;.
\end{eqnarray*}
The subscripts $n$ indicate the normalising constants, and the {\it hat} quantities are the dimensionless variables with which the code works. The expression for the plasma beta can be obtained from this normalization as
\begin{eqnarray*}
\beta=\frac{2\hat{p}}{\hat{B}^2}\;.
\end{eqnarray*}
In this paper, we will work with normalised quantities, but the {\it hat} is removed from the equations for simplicity.

The (normalised) equations governing our MHD processes are,
\begin{eqnarray}
\frac{\partial \rho}{\partial t}+\boldnabla\cdot(\rho{\bf v}) &=& 0\;,\label{n_mass}\\
\rho\frac{\partial{\bf v}}{\partial t}+\rho({\bf v}\cdot\boldnabla){\bf v} &=& -\boldnabla p + (\boldnabla\times{\bf B})\times{\bf B} + {\bf F}_{\nu}\;,\label{n_motion}\\
\frac{\partial p}{\partial t}+{\bf v}\cdot\boldnabla p &=& -\gamma p \boldnabla\cdot{\bf v}+H_{\nu}+\frac{j^2}{\sigma}\;,\label{n_energy}\\
\frac{\partial{\bf B}}{\partial t} &=& \boldnabla\times({\bf v}\times{\bf B})+\eta\nabla^2{\bf B}\;,\label{n_induction}
\end{eqnarray}
where $\eta$ is the magnetic diffusivity (which, in the normalised equations, equals the resistivity), ${\bf F}_{\nu}$ and $H_{\nu}$ are the terms for the viscous force and viscous heating, and $j^2/\sigma$ is the ohmic dissipation. The internal energy, $\epsilon$, is given by the ideal gas law, $p=\rho\epsilon(\gamma-1)$, with $\gamma=5/3$.

Finally, the time unit used in our numerical results is the FMS crossing time, $\tau_F$, i.e. the time for a FMS wave to travel from the null ($y=0$) to the top ($y=0.35$) or bottom boundary. This time is calculated as
\begin{equation}
\tau_F=\int_{y=0}^{y=0.35}\!\frac{{\rm d}y}{c_F(y)}\;,
\end{equation}
where $c_F(y)=\sqrt{v_A^2+c_s^2}$ is the local fast magnetosonic speed, and it matches the time for the initial perturbation to reach the top boundary.


\section{Initial state} \label{sec:sec2}

The starting point for our numerical experiments is of extreme importance. This state should ideally be in a perfect non-force-free equilibrium, in order to be able to compare our results with LP07. The form of such a non-force-free state has been sought by some authors in the past \cite{Fontenla93,Rastatter94,Craig05,Pontin05}, but a consistent analytical expression has still not been found. The latest attempt has been made by FF11, through the viscous, non-resistive, dynamical relaxation of a squashed potential X-point (starting from a homogeneous current density in the whole domain). The final state from this numerical relaxation is the starting point for the numerical experiments in the present paper.

The initial state was achieved numerically after a MHD evolution, consistent with energy transfer between magnetic, kinetic and internal energies. The non-zero plasma pressure makes it possible to achieve a realistic state in which the current is not located at an ideal infinitesimally thin current sheet, such as the family of solutions studied by \citet{Bungey95}. Instead, the current accumulates in a finitely thick layer which also extends along the four separatrices. The fact that this state is not in a perfect equilibrium is unavoidable, simply because of the infinite-time nature of the current singularity. Nevertheless, the time evolution of such an asymptotic regime does not cause any complications for the resistive experiments that are studied in this paper, since its time-scale is much slower than those of the processes discussed here.

The final state of the numerical relaxation found in FF11 is described as a ``quasi-static'' equilibrium in which the Lorentz and the pressure forces balance each other throughout the domain, save at the current layer. As a MHS equilibrium, this state must obey the Grad-Shafranov equation, which, for 2D symmetries, relates the plasma pressure, flux function and current density, as follows
\begin{equation}
\frac{{\rm d}p(A_z)}{{\rm d}A_z} = -j_z(A_z)\;. \label{gradshaf}
\end{equation}

Equilibrium is achieved everywhere except locally at the null and along the separatrices. After the viscous relaxation has finished, the system enters a slow asymptotic regime towards an infinite time singularity. For further details, see FF11.

\begin{figure}[t]
  \centering
  \includegraphics[scale=0.30]{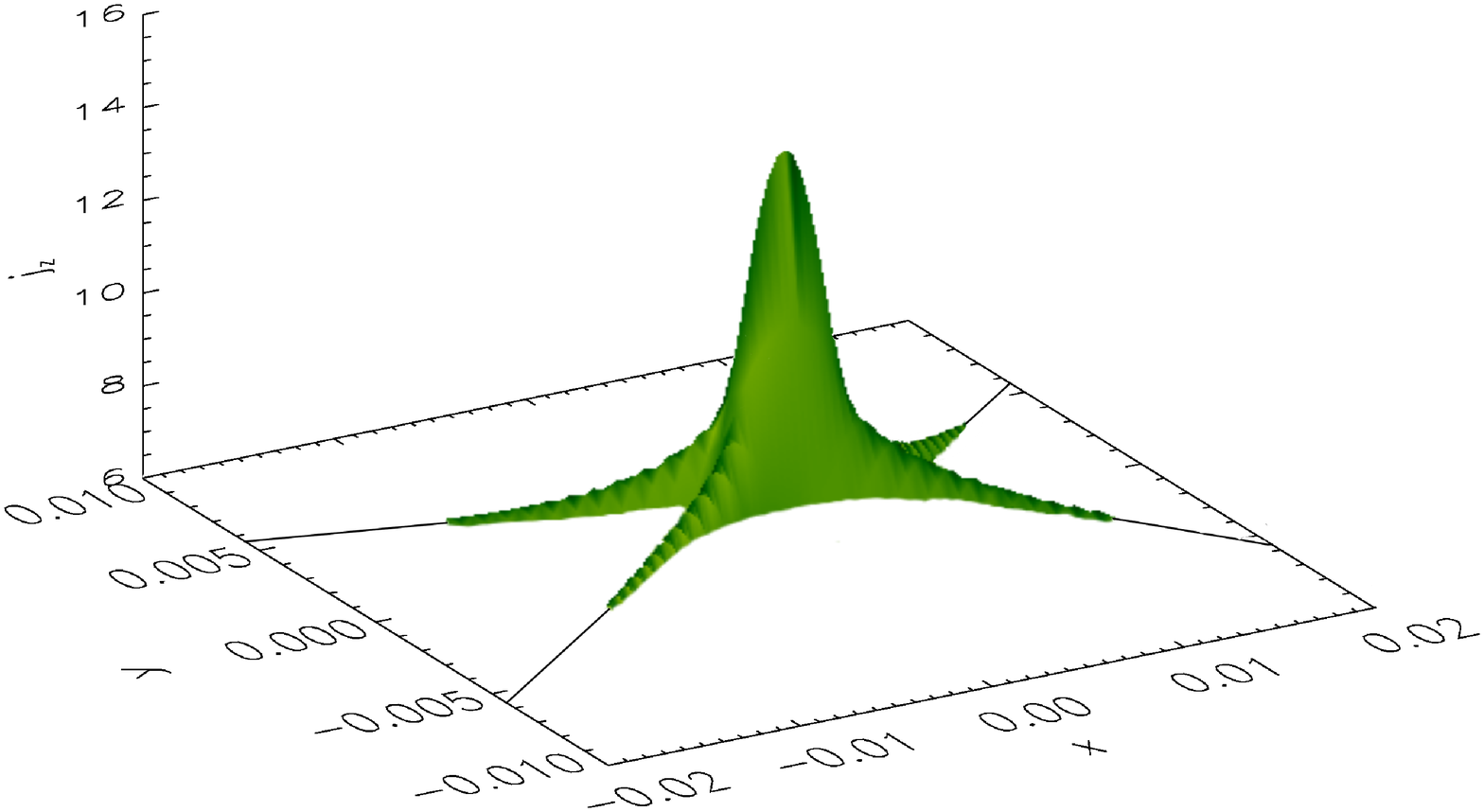}
  \caption{Current density surface within the initial diffusion region, in a close up of the null point. Magnetic field separatrices are shown as black lines. Note, the current extended along the separatrices is only just over $j_{crit}=6$.}
  \label{fig:initial_dr}
  \vspace{0.3cm}
  \includegraphics[scale=0.30]{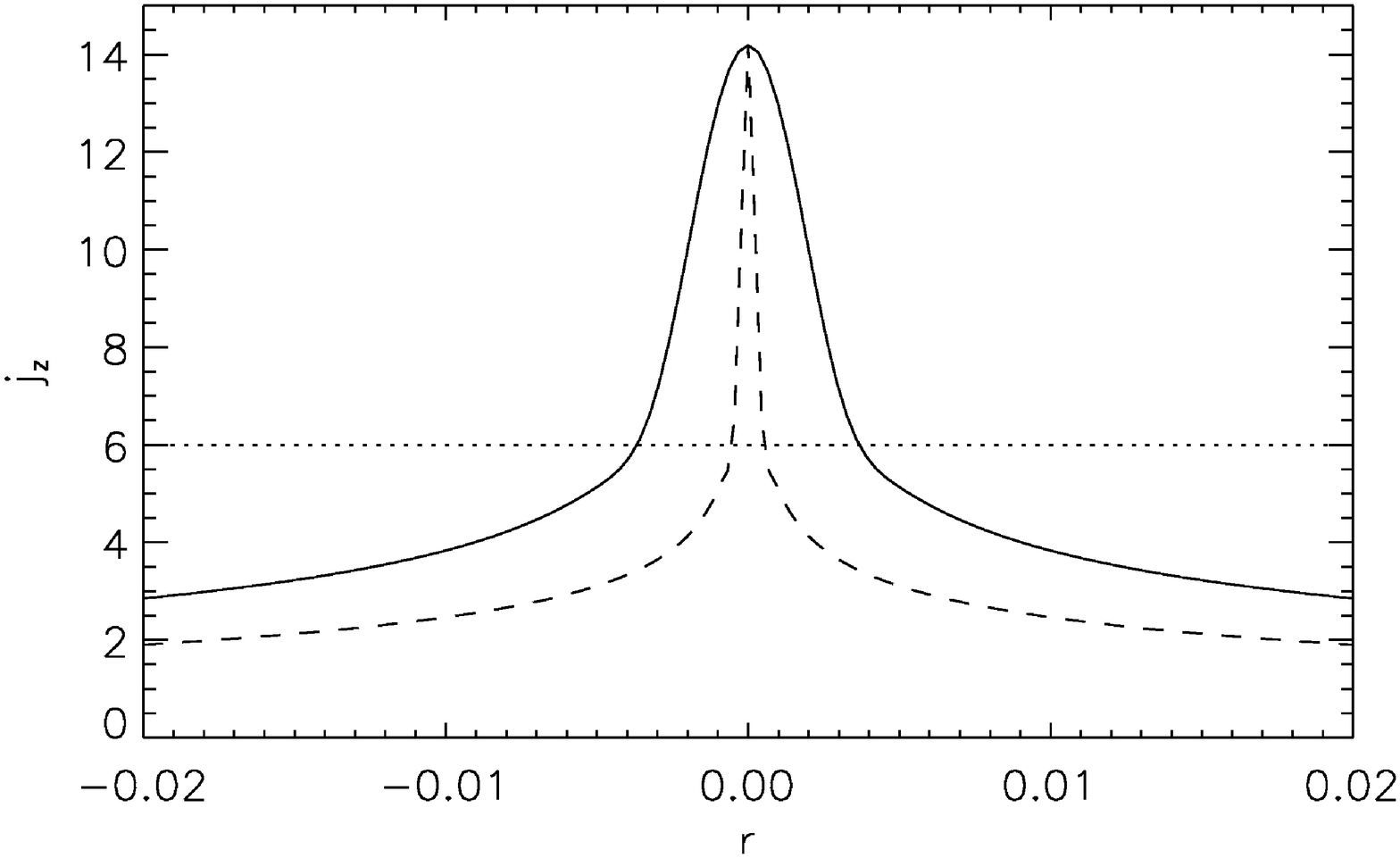}
  \caption{Cuts of current density in the $x$-direction (solid) and $y$-direction (dashed) across the central current layer. The radial coordinate $r$ corresponds to $x$ for the solid line, and $y$ for the dashed line. The dotted horizontal line marks the level of $j_{crit}$.}
  \label{fig:initial_cs}
\end{figure}

Fig. \ref{fig:initial} shows the final state from FF11 which is the starting point for our numerical experiments. The current density (Fig. \ref{fig:initial}a) is localised within a small region about the null and faintly extends along the four separatrices. The plasma pressure (Fig. \ref{fig:initial}b) has a positive gradient in the regions inside the cusp (left and right), and a negative gradient outside the cusp (top and bottom). The initial background values of the current density and the plasma pressure for this particular experiment are $j_0=1.04$ and $p_0=0.375$, and the viscosity is isotropic and has a value of $\nu=0.0001$. The resolution of the numerical grid is $3072\times 3072$, which uniformly covers the domain extending from $-0.5$ to $0.5$ in the $x$-direction and $-0.35$ to $0.35$ in the $y$-direction. Magnetic field lines are line-tied at the four boundaries and all components of the velocity are set to zero on the boundaries. The other quantities have their derivatives perpendicular to each of the boundaries set to zero. Both total energy and total mass are conserved in the experiments, to within numerical error.

Starting with the FF11 set-up (Fig. \ref{fig:initial}), we apply an anomalous (non uniform) resistivity that vanishes below a critical value of the current density, $j_{crit}$, and is constant above it,
\begin{equation}
  \eta =
  \left\{
  \begin{array}{ll}
    0  & \mbox{if } j_z < j_{crit}\;, \\
    \eta_0 & \mbox{if } j_z \ge j_{crit}\;.
  \end{array}
  \right.
\end{equation}
This approach allows diffusion only about the null, where the current density is large. This is known to be a crucial parameter for fast reconnection regimes and solar flares \cite{Yokoyama94,Ugai03,Ugai05,Uzdensky05}, but also for coronal heating from small-scale sources \cite{Roussev02,Krasnoselskikh02}. Here, we have chosen $j_{crit}=6$ and $\eta_0=0.0008$. The estimated numerical diffusivity, due to the finite grid resolution, is $0.0001$. In the absence of an enhanced diffusivity, $\eta_0$, the current layer would keep evolving slowly in time becoming shorter and thinner (as seen in Fig. 9 in FF11).

It is worth noting that the initial structure of the current density extends along the separatrices very faintly, as shown in Fig. \ref{fig:initial_dr}, but the main concentration of current is still at the central current layer. In Fig. \ref{fig:initial_cs}, we show horizontal and vertical cuts through the null point, of the initial current density distribution. Note, how the central current layer is elongated in the $x$-direction, but is much thinner in the $y$-direction.


\section{Results} \label{sec:sec3}


\subsection{Diffusive phase} \label{sec:sec3.1}

Following the approach of LP07, we first look at the initial phase where the current density peak is diffused by the resistive terms, to the point where it reaches the value of $j_{crit}$. This diffusive phase occurs within the first few time-steps of the simulation and contains all the energy conversion due to the ohmic dissipation itself.

\begin{figure*}[t]
  \centering
  \includegraphics[scale=0.62]{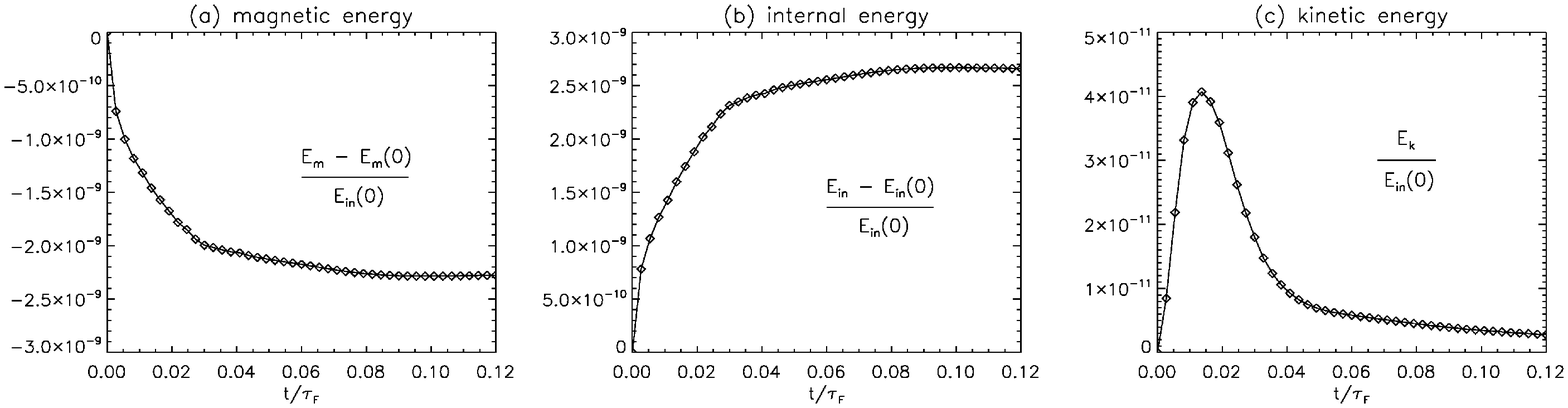}
  \caption{Time evolution of a) magnetic, b) internal and c) kinetic energy, integrated over the whole domain, for the first time-steps of the simulation. The first two have been normalised to the relative change from their non-zero initial values. The initial values of the magnetic and internal energies are, respectively, $0.09$ and $0.40$.}
  \label{fig:energies}
  \vspace{0.5cm}
  \includegraphics[scale=0.62]{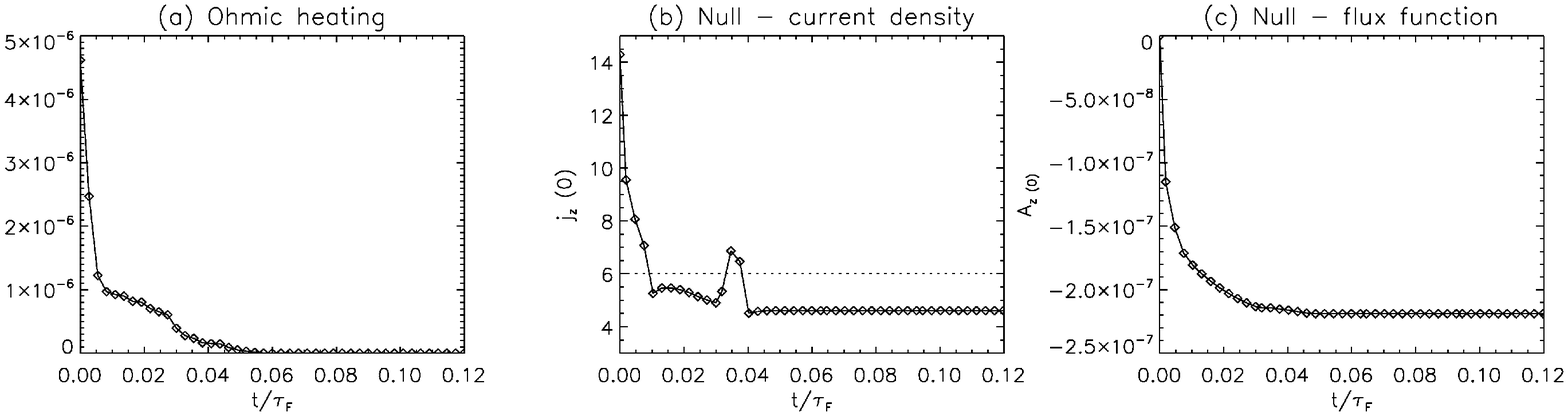}
  \caption{Time evolution of a) the ohmic heating integrated over the domain, b) the current density at the location of the null (the dotted line indicates the value of $j_{crit}$) and c) the flux function at the location of the null, for the same time interval as in Fig. \ref{fig:energies}.}
  \label{fig:nulldiag}
\end{figure*}

In Fig. \ref{fig:energies}, we show the changes in magnetic, internal and kinetic energies, normalised to the value of the initial internal energy. The kinetic energy is zero in the initial state. From these plots, we can see the sudden decrease of magnetic energy (Fig. \ref{fig:energies}a), due to the magnetic reconnection occurring at the null, which is directly transferred into both internal (Fig. \ref{fig:energies}b), and kinetic energy (Fig. \ref{fig:energies}c). The relative changes of energy are very small. This is due to the very narrow initial diffusion region (see Fig. \ref{fig:initial_dr} and \ref{fig:initial_cs}), which only allows a small amount of magnetic flux to be reconnected. The energies remain constant after the time $t/t_F$ shown in the plots.

In Fig. \ref{fig:nulldiag}, we show the ohmic heating integrated over the whole domain, together with the behaviour of the current density, $j_z$, and the flux function, $A_z$, at the location of the null. The ohmic heating (Fig. \ref{fig:nulldiag}a), which is proportional to the absolute change of $j_z^2$, is non-zero from the first moment, at which reconnection starts occurring, and then decreases rapidly. At a time of about $t\sim 0.05\tau_F$, the ohmic heating vanishes, indicating that the current density is now below $j_{crit}$ everywhere in the domain, and hence, reconnection has stopped.

The time evolution of the null current density (Fig. \ref{fig:nulldiag}b) shows a complicated pattern that is difficult to interpret. Initially, the current at the null decreases rapidly until $j_z<j_{crit}$ (at $t\sim 0.01\tau_F$). Note that, this occurs just before the kinetic energy reaches its maximum, and just before the ohmic heating flattens. This current density overshoot (the current going well below the level of $j_{crit}$) may be a consequence of the small scale advection, outwards from the null, within the expanded diffusion region, right after the sudden decrease of the peak current density, which spreads out the current density lowering it below $j_{crit}$. At time $t\sim 0.03\tau_F$, the null current density increases significantly (even above the value of $j_{crit}$) and it then quickly goes back down.  A possible cause for this effect is discussed below. Soon after, it stabilises to a value of $j_z\sim 4.5$, at roughly the time when the ohmic heating stops.

In Fig. \ref{fig:nulldiag}c, we show the time evolution of the flux function at the null point (initially zero for convenience), normalised to the value of $A_z$ at the middle ($x=0$) of the upper boundary (i.e. to the range of $A_z$ available to reconnect, that is 0.5). The change of the flux function is negative, because the reconnected field lines have a negative $A_z$. The rate of change of $A_z$ decreases at $t\sim 0.03\tau_F$, in agreement with the null current density rise, and then stops at the time when ohmic heating ceases, so $A_z$ remains fixed here-after.

\begin{figure}[t]
  \centering
  \includegraphics[scale=0.3]{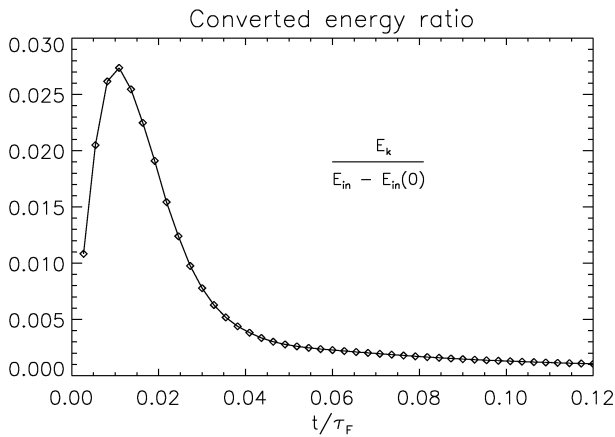}
  \caption{Time evolution of the ratio of the change in kinetic energy to the change in internal energy, for the same time interval as in Fig. \ref{fig:energies}.}
  \label{fig:ratio}
\end{figure}

We now investigate further the time evolution of the energies of the system, to determine the nature of the energy into which the majority of the transferred magnetic energy is converted. Fig. \ref{fig:ratio} shows the ratio of the change in kinetic energy to the change in internal energy. It is clear that the vast majority of the converted magnetic energy during the reconnection process goes directly into internal energy, while the energy that is transferred into kinetic is minimal. This is the first main result that differs dramatically from LP07. Our initial state is not a potential equilibrium with a current sheet, but is a non-force free state in which there exists a background non-zero current density everywhere, together with a plasma pressure gradient which is large compared to the magnitude of the effects driven by the reconnection process. This contrasts with the zero-beta treatment of LP07 which did not include internal energy.

The consequences of this result are that the amplitude of the propagating wave is extremely small. Changes in current density and pressure are no more than about $10^{-6}$ times their initial values. Thus, for the analysis of the propagating wave, we have had to consider the differences of quantities from their initial non-uniform distributions, rather than their absolute magnitudes.

We now return to the plots of the time evolution of the null current density (Fig. \ref{fig:nulldiag}b). A possible cause for the current density bump observed at $t\sim 0.03\tau_F$ is the following. During the first 3 or 4 time-steps, the current density peak is reduced following a diffusive solution, until the null current density has been decreased to a level below $j_{crit}$ (i.e. at $t\sim 0.03\tau_F$). During this period, a small amount of magnetic energy is converted into internal energy via ohmic heating, and a tiny amount of it is converted into kinetic energy, accelerating the plasma within the diffusion region. By the end of this diffusive phase, a fast magnetosonic wave is launched from the edge of the diffusion region. But instead of a single pulse outwards from the null, as described in LP07, a pair of FMS pulses of opposite sign, directed outwards from the diffusion region edge and inwards towards the null point, is launched. The closest distance between the edge of the diffusion region and the null appears in a vertical cut through the null point (see Fig. \ref{fig:initial_dr}). This width is approximately $0.00055$. Also, the speed of the FMS wave in the region near the null can be approximated as the sound speed of the plasma in that region (since the magnitude of the magnetic field is negligible near the null), which in our experiment is $c_s\sim0.9$. This gives a time for a magnetosonic pulse to travel from the closest edge of the diffusion region to the null of $t\sim 0.02\tau_F$. Added to the time when the current density falls to the value of $j_{crit}$, gives a value of $t\sim 0.03\tau_F$, which is the time of the observed current density bump.

\begin{figure*}[t]
  \centering
  \includegraphics[scale=0.21]{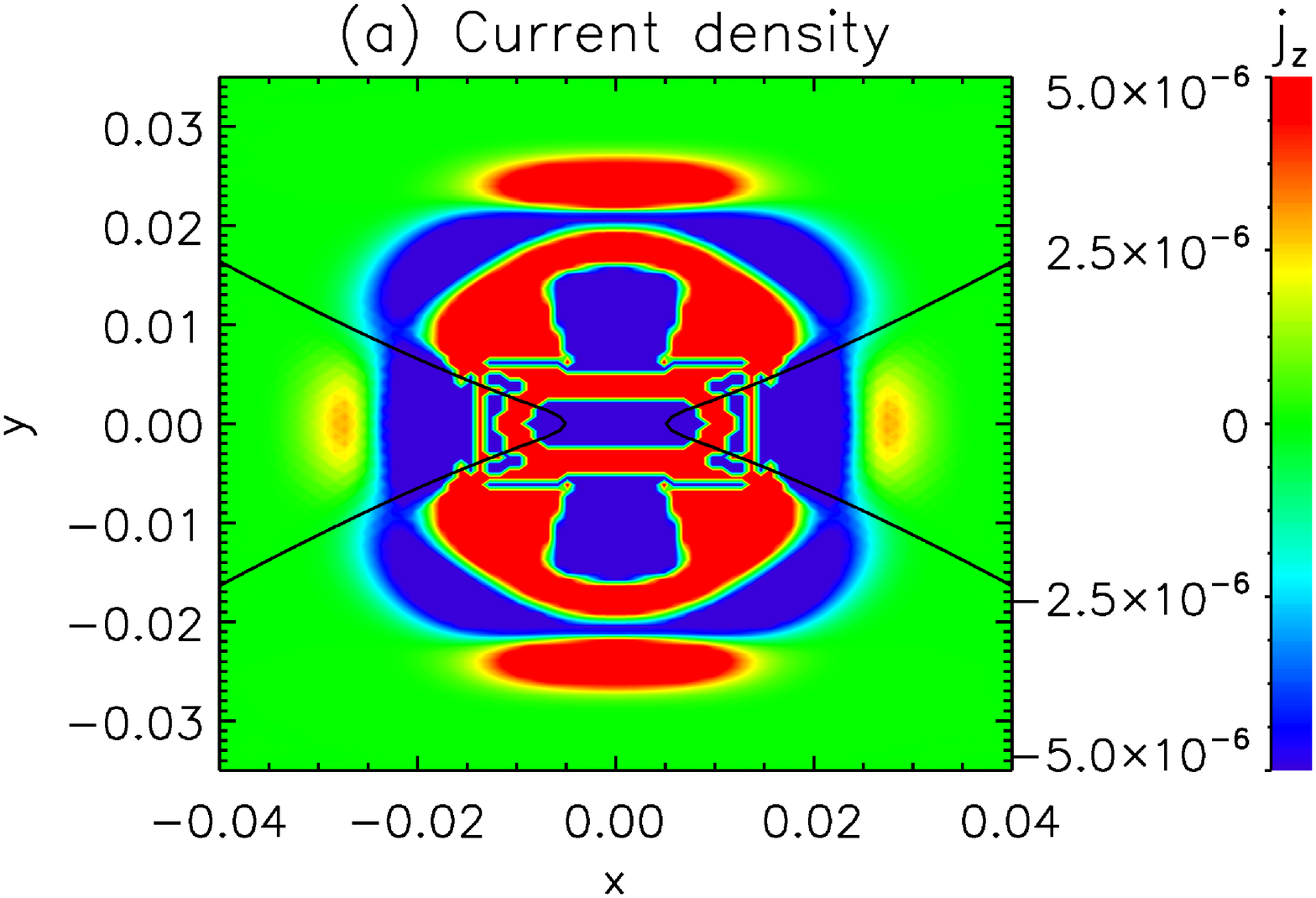}
  \includegraphics[scale=0.21]{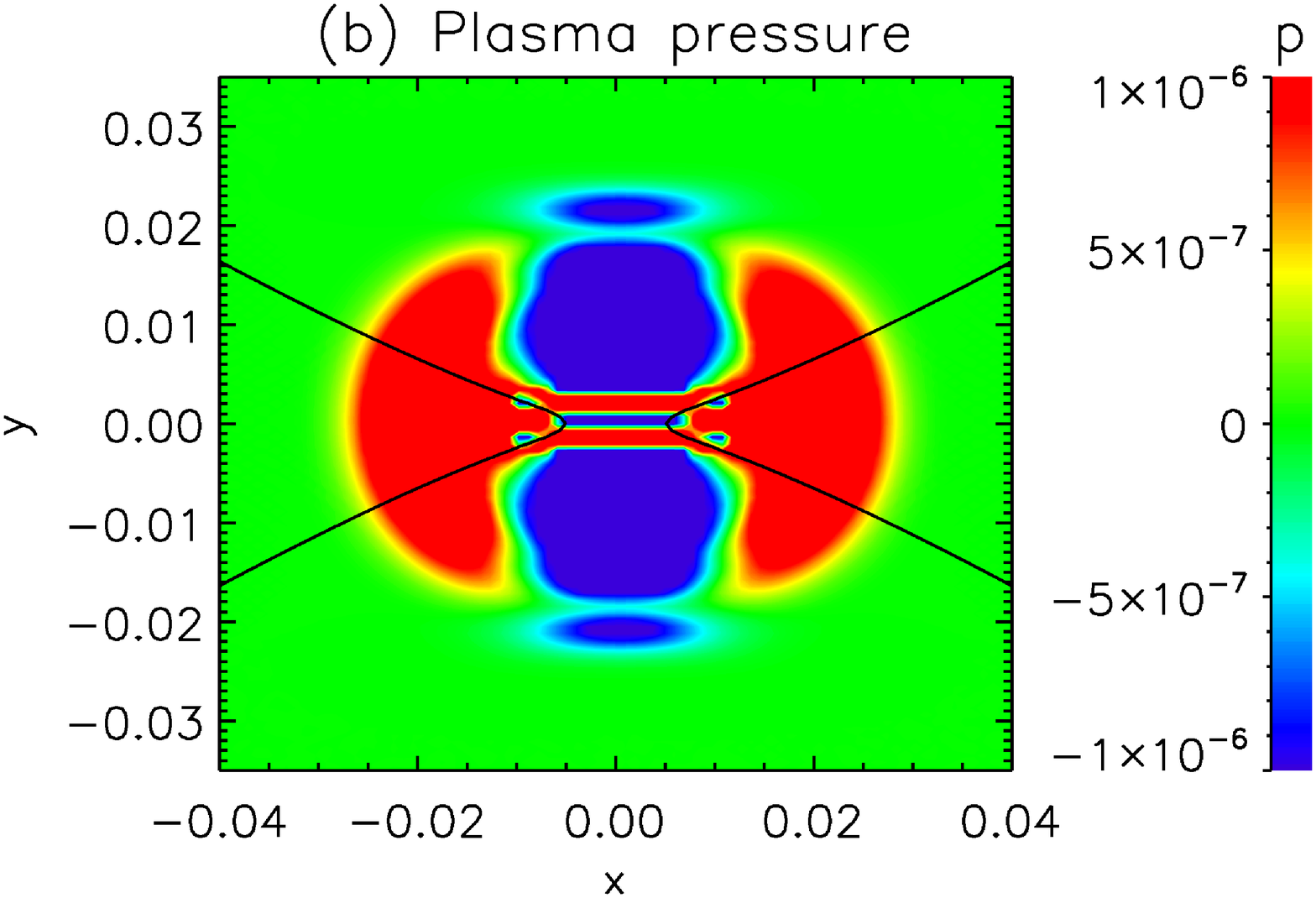}
  \includegraphics[scale=0.21]{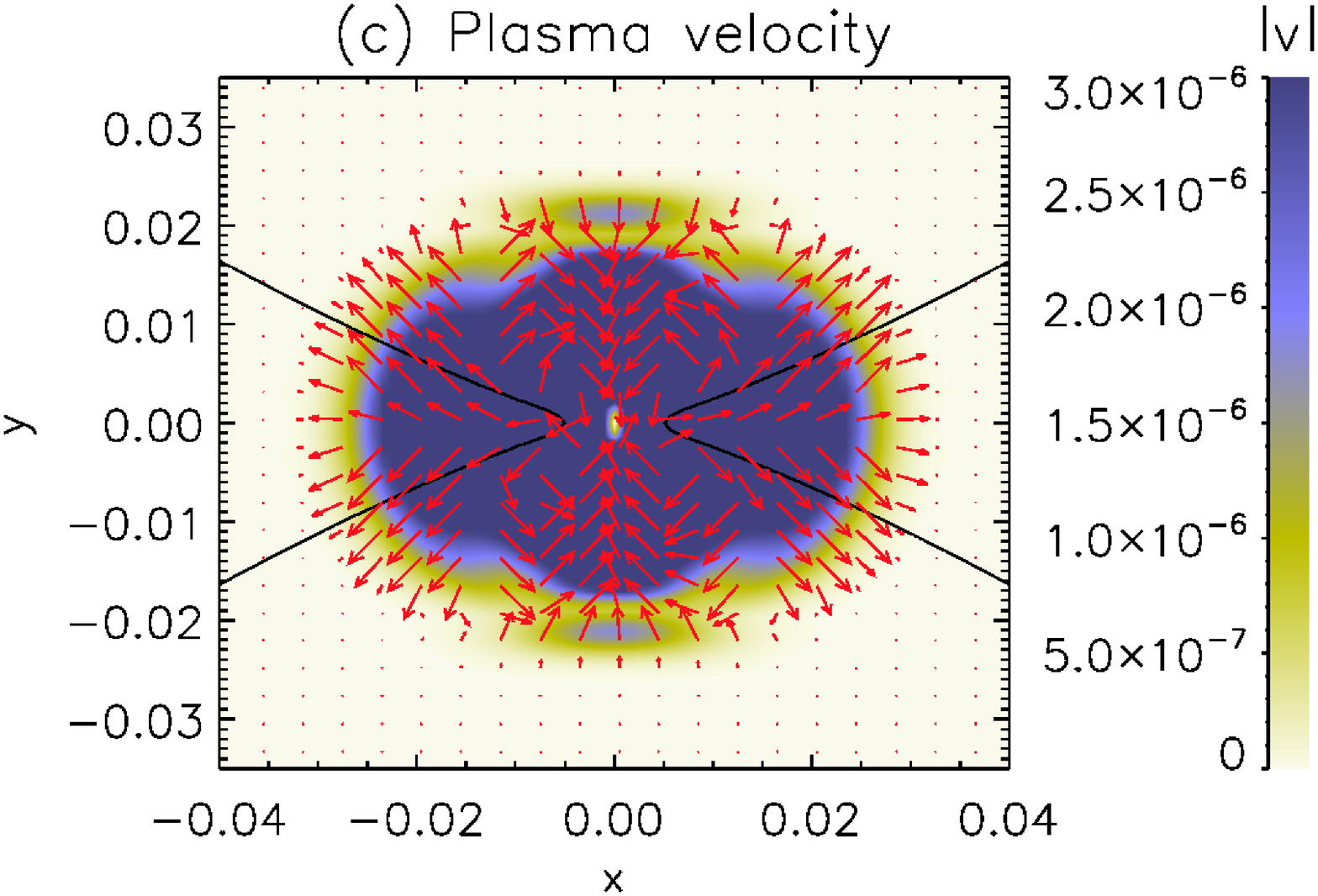}
  \caption{Two-dimensional contour plots of (a) electric current density perturbation (b) plasma pressure  perturbation and (c) magnitude of the velocity field, for $t=0.06\tau_F$. Solid black lines show the magnetic field lines with$A_z=0$, i.e. the separatrices for the initial state (hence, a sign for reconnection to have occurred). In (c), red arrows show the direction of the flow at each point.}
  \label{fig:2Dmaps_ini}
\end{figure*}

\begin{figure}[t]
  \centering
  \includegraphics[scale=0.28]{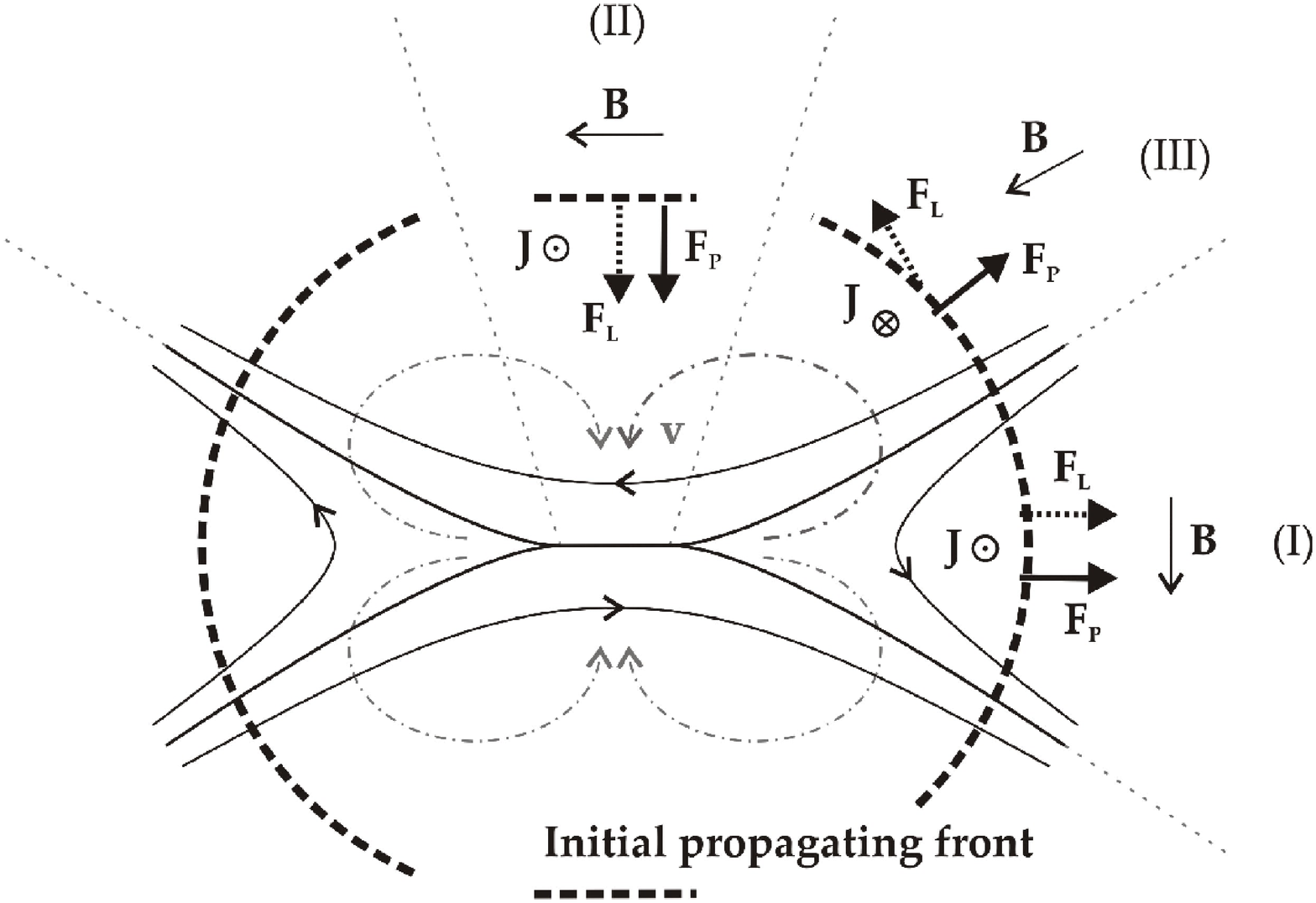}
  \caption{Sketch of the different forces that are carried by the magnetic and plasma perturbations, at any moment before the pulse has left the diffusion region well behind. The dashed thick line is the propagating wave front. Here, we show the current density vector corresponding to each of the pulses (positive current points outside of the 2D plane), the direction of the background magnetic field vector, and the directions of the plasma and Lorentz forces, ${\bf F_p}$ and ${\bf F_L}$ respectively. }
  \label{fig:diagram1}
\end{figure}

In other words, if a pair of fast waves is launched from the edge of the diffusion region in opposite directions at the moment when the diffusion stops, i.e. when $j_z<j_{crit}$, then, one of these pulses would reach the null, causing an increase in the current density, and acting against reconnection for that short period of time. This effect can be seen clearly in Fig. \ref{fig:nulldiag}b and causes a minor perturbation in the flux function at the null, as seen in Fig. \ref{fig:nulldiag}c. Of course, the same thing would happen from the below edges of the diffusion region, and hence, these two pulses would travel through and follow the opposite partners outwards from the null.

Now that the initial diffusive phase has been described, we analyse the subsequent phase where a FMS wave is launched radially from the null point. The signature of this wave is seen as a small perturbation to the current density, the plasma pressure and the plasma flow velocity.


\subsection{Wave initiation}

Immediately after the short diffusive phase in which the current is reduced to below the level of $j_{crit}$, a FMS wave is launched in all directions, in accordance with LP07. As already discussed, due to the small amplitude of our perturbation, from now on we show all quantities subtracted from the initial distributions. Fig. \ref{fig:2Dmaps_ini} shows the current density and plasma pressure perturbations, and the velocity vector field, for a time shortly after the initiation of the wave ($t=0.06\tau_F$). An important fact to note is that the dominant mode in the magnetic field perturbation is $m=4$, as seen from the current density map (Fig. \ref{fig:2Dmaps_ini}a). We first focus on the velocity vector field (Fig. \ref{fig:2Dmaps_ini}c) and the plasma pressure pulse (Fig. \ref{fig:2Dmaps_ini}b).

The velocity field pattern (inflow above and below the null, and outflow to the left and right) is caused by the passage of the FMS wave communicating the occurrence of the sudden reconnection that has occurred at the thin diffusion region. The plasma inflows leave a pressure deficit behind, while the plasma outflows cause a pressure enhancement in front. This explains the different signs of the plasma pressure perturbation (Fig. \ref{fig:2Dmaps_ini}b). A pair of positive semi-circular pulses travel outwards from the left and right edges of the diffusion region, and a pair of negative planar pulses (remember we are dealing with differences here, not with absolute quantities) travel above and below the null. The positive pressure pulses correspond to plasma outflows, while the negative pressure pulses are associated with plasma inflows. Also, the shapes of the positive and negative pulses are defined by the initial shape of the diffusion region at the time where the FMS wave is launched. This is about 5 times longer than wider, and its short and faint extensions along the 4 separatrices are one order of magnitude smaller than in the central region (see Fig. \ref{fig:initial_dr}) and so, their contribution may be completely neglected. Hence, the negative plasma pressure pulse is initiated in a planar way, while the positive pulses are initiated at the left and right edges of the diffusion region, showing a circular propagation centered at those vertices.

\begin{figure*}[t]
  \centering
  \includegraphics[scale=0.21]{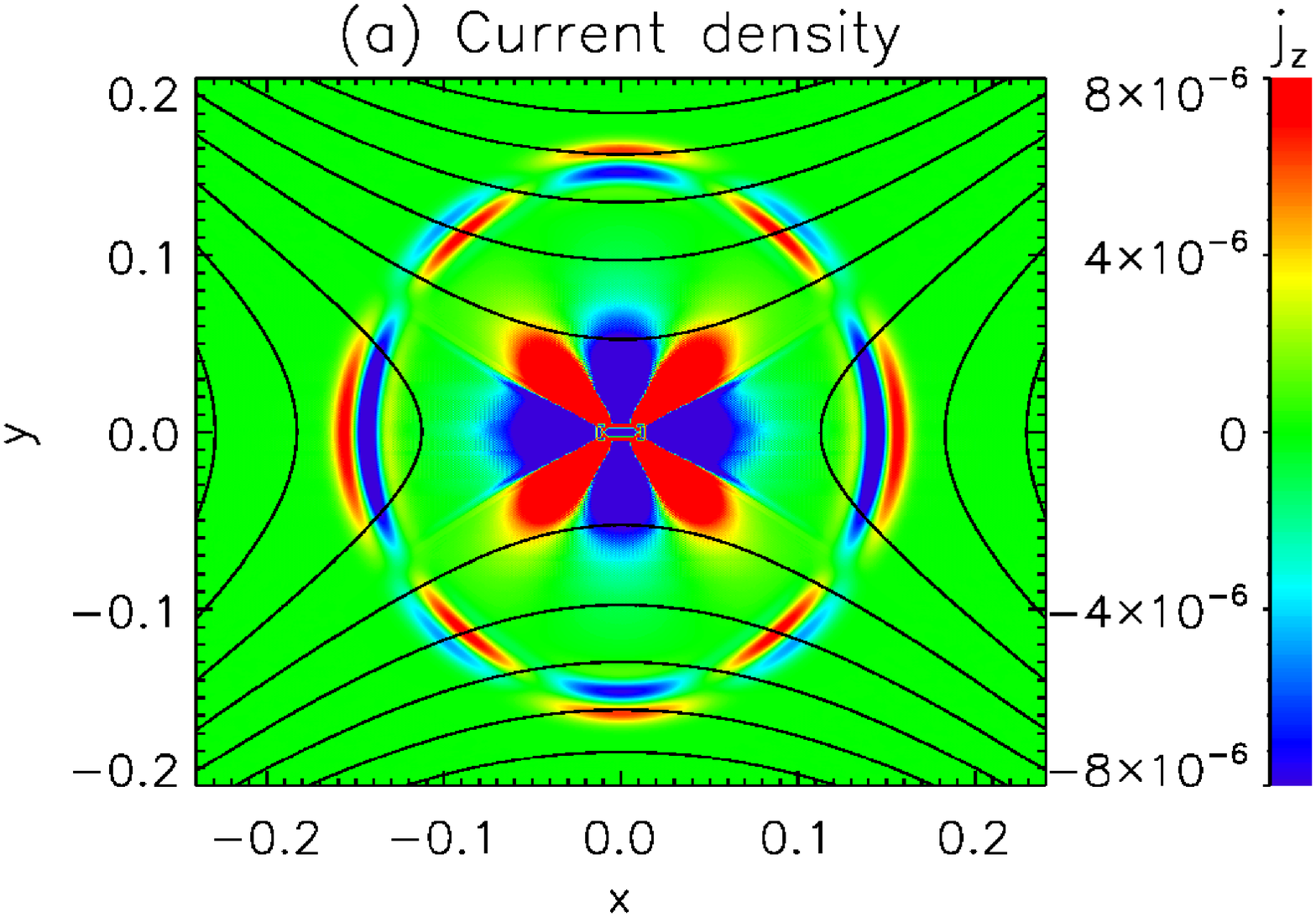}
  \includegraphics[scale=0.21]{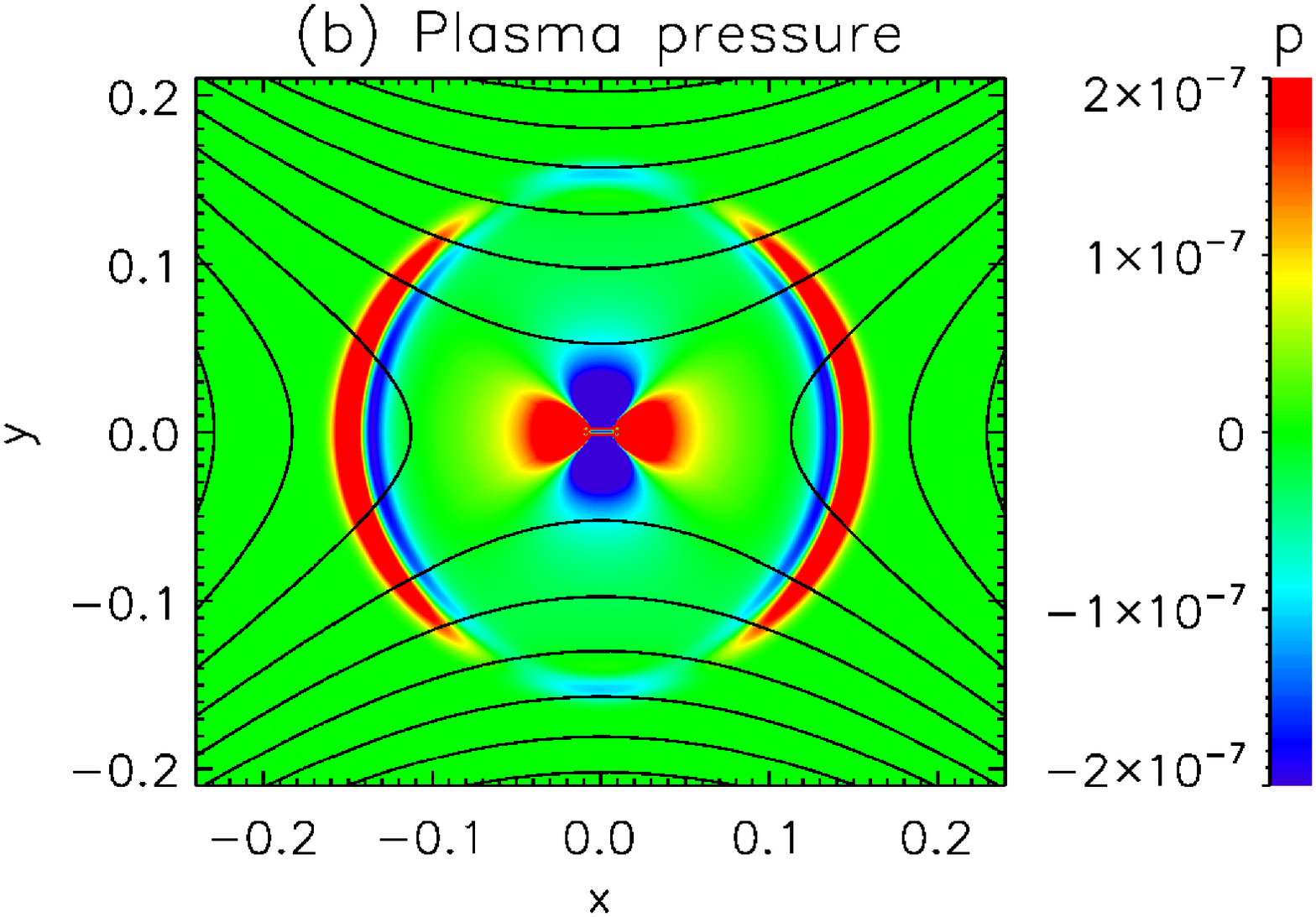}
  \includegraphics[scale=0.21]{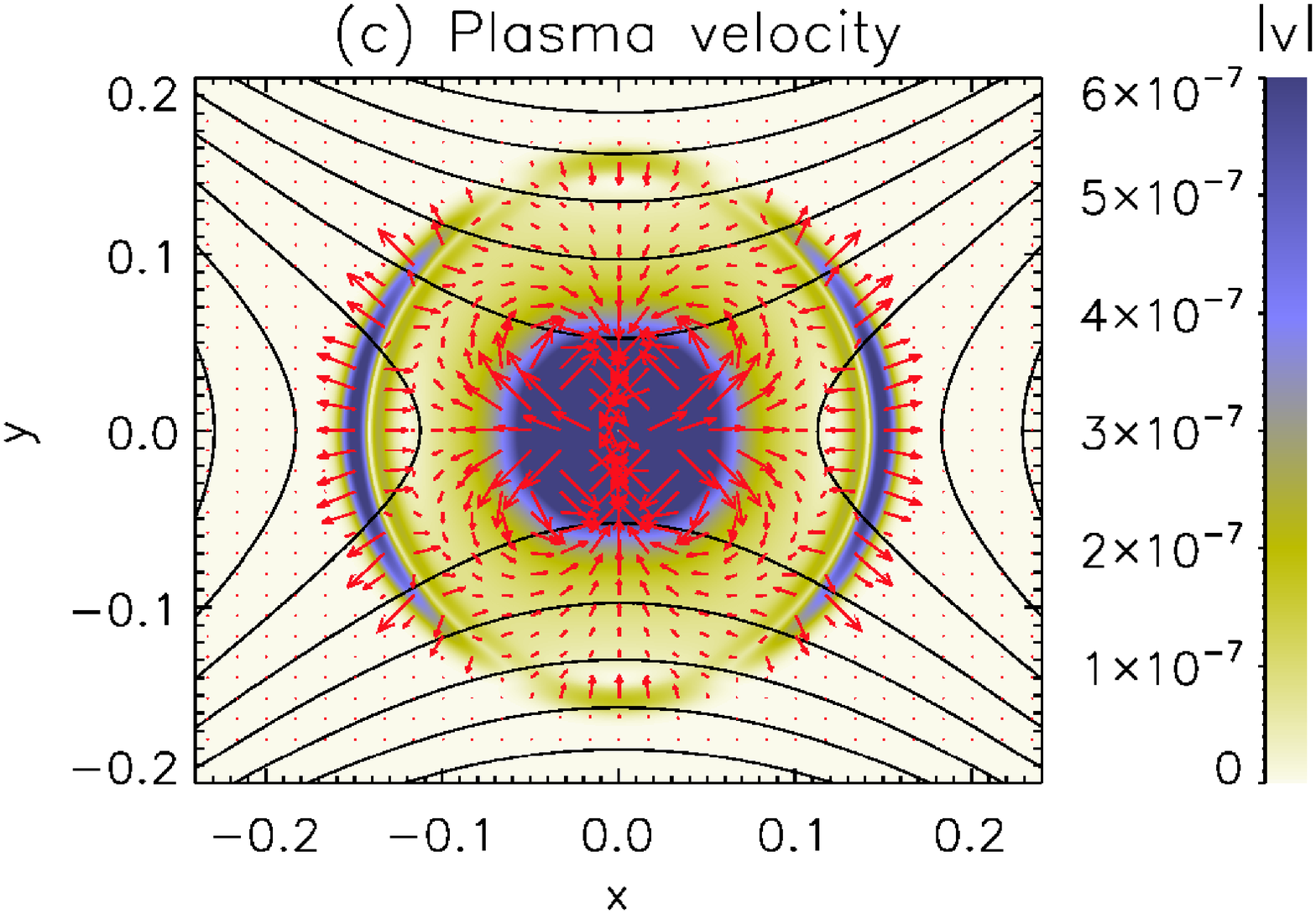}
  \caption{Two-dimensional contour plots of (a) electric current density perturbation (b) plasma pressure perturbation and (c) magnitude of the velocity field, for $t=0.49\tau_F$. Solid black lines are the magnetic field lines. In (c), red arrows show the direction of the flow at each point.}
  \label{fig:2Dmaps}
\end{figure*}

\begin{figure}[t]
  \centering
  \includegraphics[scale=0.28]{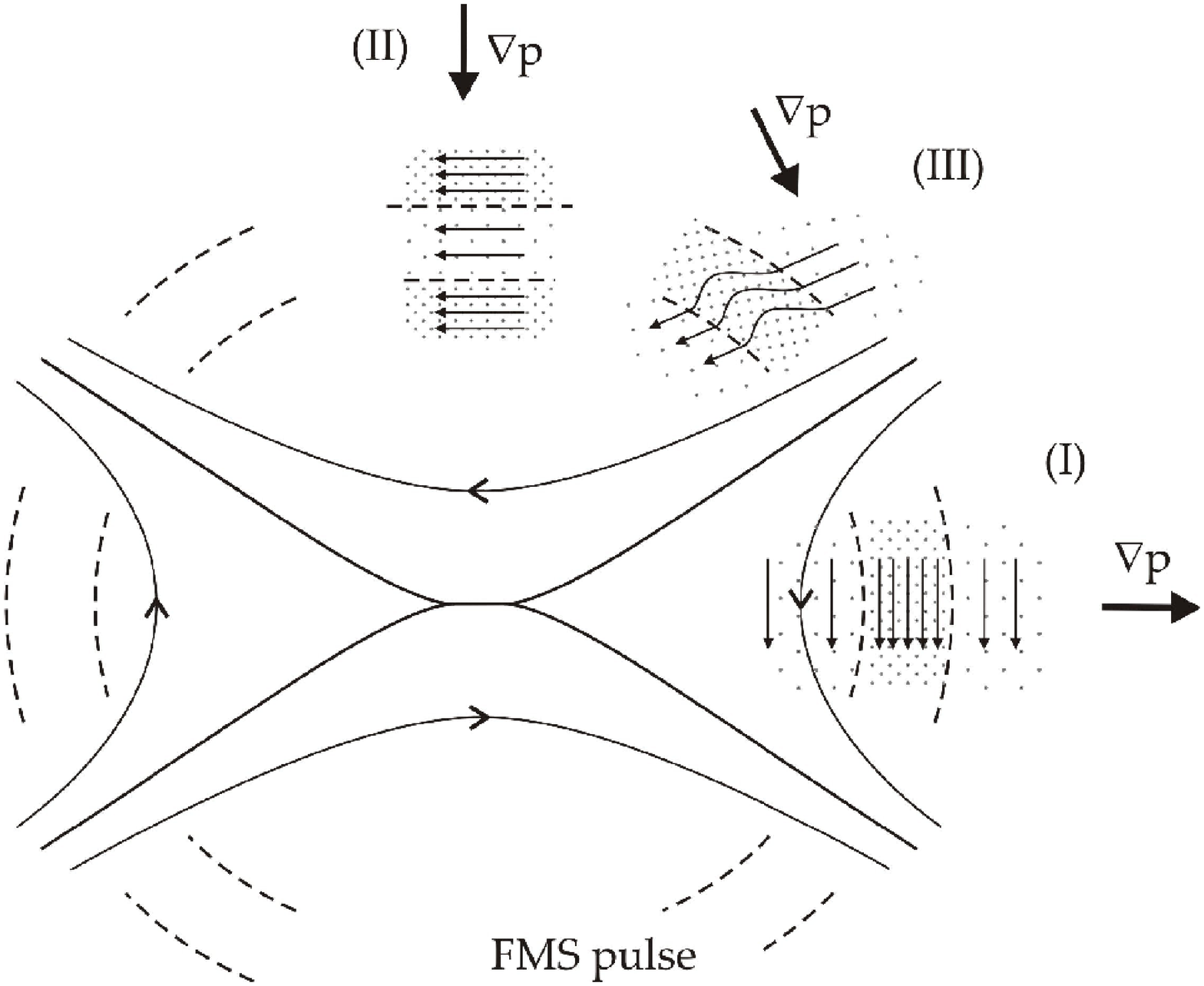}
  \caption{Sketch of the magnetic field perturbation and the leading pressure front perturbation for the three same regions as in Fig. \ref{fig:diagram1}. The background plasma pressure gradient is indicated for each region.}
  \label{fig:diagram2}
\end{figure}

In order to understand the structure of the current density pulse (Fig. \ref{fig:2Dmaps_ini}a), we evaluate the forces that the FMS wave is producing, and we compare them with the velocity flow pattern. These forces are sketched in Fig. \ref{fig:diagram1} for three different zones: region (I), to the left and right, where both the pressure pulse and the leading current pulse are positive; region (II), top and bottom, where the pressure pulse is negative and the leading current pulse is positive; and region (III), the diagonals, where the pressure pulse is positive but the leading current pulse is negative. The pressure enhancements in regions (I) and (III) create an outwards pressure force, ${\bf F_p}$, while the pressure deficit in region (II) creates an inwards pressure force. These forces are consistent with reconnection. Also, the cross product of the leading current density perturbation and the background magnetic field, causes a different Lorentz force, ${\bf F_L}$, in each region, which matches the motion of the plasma due to the reconnection flows. In conclusion, the structure of both the plasma pressure and the current density propagating pulses are such that they create the pertinent forces consistent with the reconnection flows.


\subsection{Wave propagation}

As the FMS wave escapes further from the central diffusion region, it leaves behind a region of magnetic field and plasma that has returned to the initial equilibrium state. This is shown in Fig. \ref{fig:2Dmaps}. Here, we see the current density and plasma pressure perturbations, and the velocity vector field, for a time when the propagating wave is far from the region close to the null ($t=0.49\tau_F$). The $m=4$ mode is again clearly seen in the current density perturbation (Fig. \ref{fig:2Dmaps}a), and the same structure described above remains during the propagation of the wave. Note, the propagating current density pulse always carries both a negative and a positive current.

As for the plasma pressure pulse (Fig. \ref{fig:2Dmaps}b), we can see the propagation of two faint negative pulses travelling vertically at the top and bottom, and the two curved positive fronts to the right and left of the null. When the positive pulse leaves the region about the null, a pressure deficit appears following behind. By looking at the velocity vector field (Fig. \ref{fig:2Dmaps}c), we see that this corresponds to a slow encroachment towards the origin, and is the result of advection from the wave interacting with diffusion (see LP07 for more details).

The waves observed in the previous plots expand with the local fast speed. However, due to the asymmetry of our domain, the pulse is not perfectly concentric, but expands slightly faster in the vertical direction.

The physical perturbation that the FMS wave causes in the magnetic field and the plasma is sketched in Fig. \ref{fig:diagram2}, for the same three regions as described in Fig. \ref{fig:diagram1}. In region (I), the FMS wave compresses both the plasma and the magnetic field (the following decompression of the plasma due to the flow encroachment is not shown in the Figure for simplicity). In region (II), it causes the opposite behaviour, expanding both the plasma and the magnetic field. And in region (III), the FMS wave bends the magnetic fields towards the direction of the reconnection flow, and compresses the plasma.

\begin{figure}[t]
  \centering
  \includegraphics[scale=0.38]{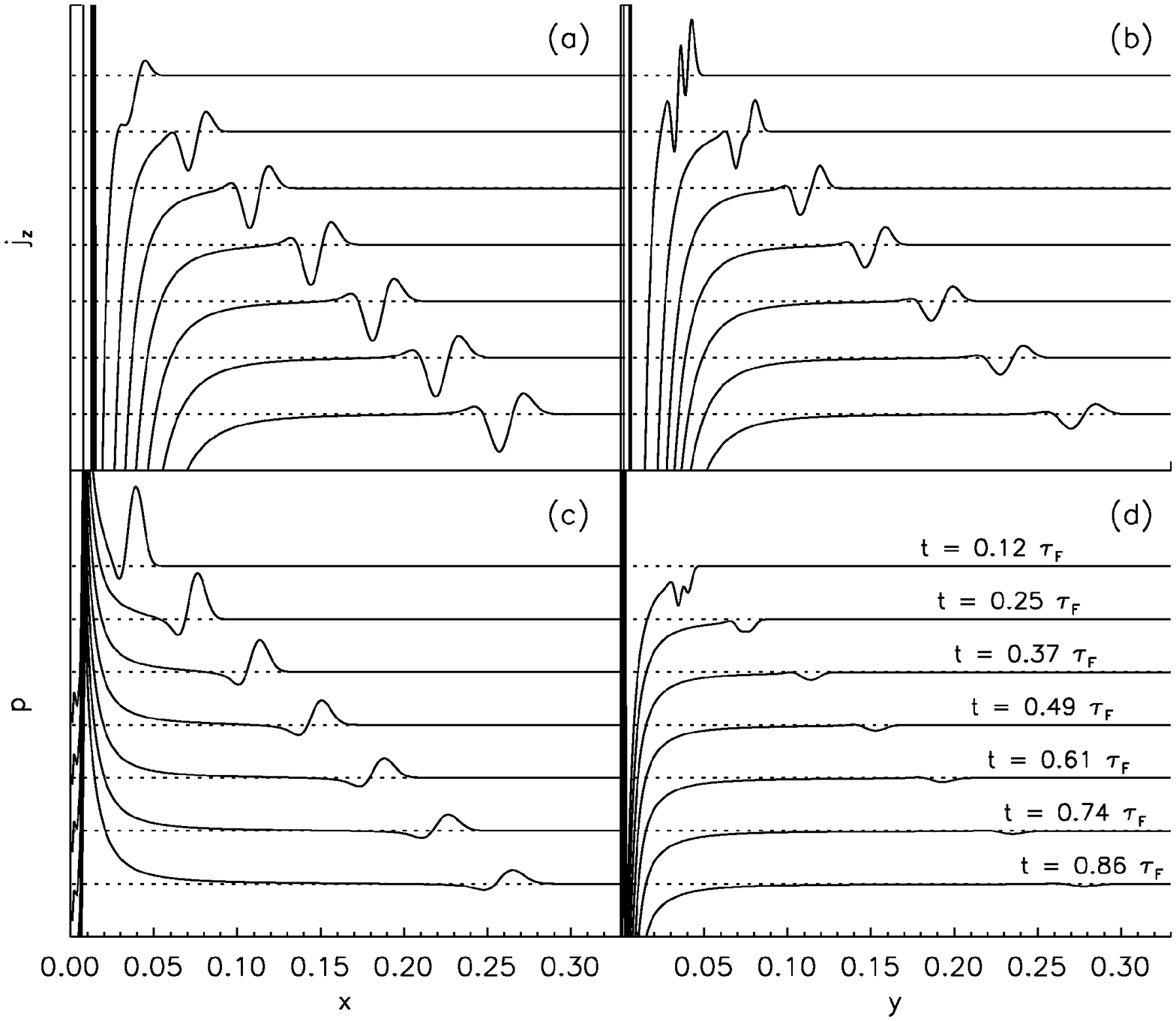}
  \caption{One-dimensional plots of current density (above) and plasma pressure (below) in a horizontal (left) and vertical (right) cut in the middle of the box, through the null, for seven different times during the propagating wave phase. Curves are displaced in the $y$-axis, and the dotted lines mark the zero for each of them. The times for all cuts are given on (d).}
  \label{fig:rtjzp}
  \includegraphics[scale=0.31]{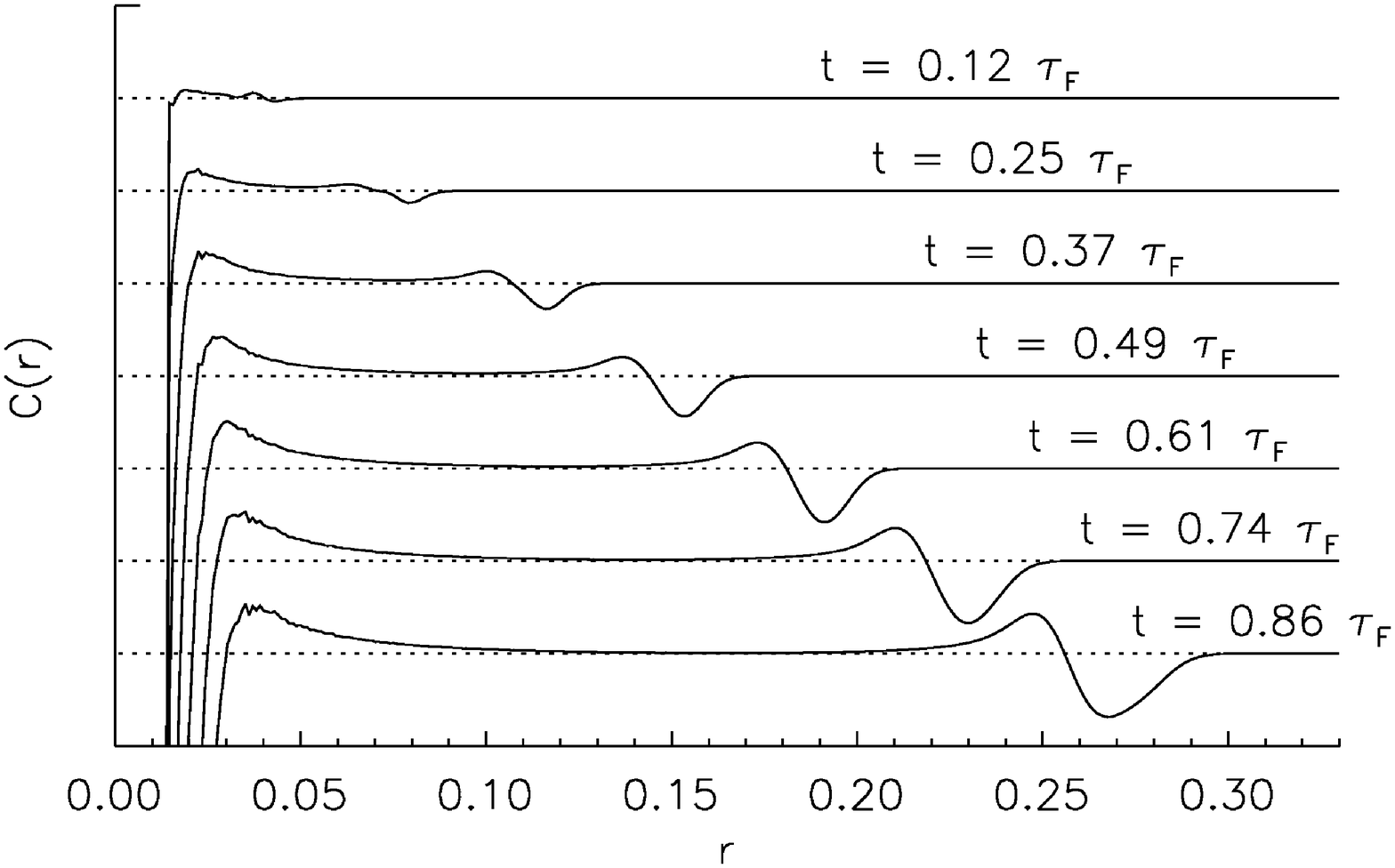}
  \caption{Plot of the quantity $C(r)$, from Eq. (\ref{circle}), at the same times as in Fig. \ref{fig:rtjzp}, as a function of radius. Curves are displaced in the $y$-axis, and the dotted lines mark the zero for each of them.}
  \label{fig:circle}
\end{figure}

Over all, what Figures \ref{fig:2Dmaps_ini} to \ref{fig:diagram2} show is that the nature of the wave propagation in each region is different, according to how the forces act in each of them. What is more, the reconnection does not produce one single pulse, but four (two quasi-planar pulses at the top and bottom of the layer, and two semicircular pulses at the right and left edges). Thus a dominating $m=0$ radial mode for the magnetic field perturbation could not exist, because the magnetic field is not radially symmetric, and neither are the forces caused by the reconnection.

In order to have a closer look at the propagation of the wave, we look at its profile in different cuts at various times during the numerical simulation. Fig. \ref{fig:rtjzp} shows horizontal and vertical cuts in the domain, from the null point to the boundaries, of both the current density and the plasma pressure, for seven times throughout the numerical experiment, before the pulse reaches the upper and lower boundary ($t=\tau_F$). In the horizontal cut of current density (Fig. \ref{fig:rtjzp}a) we can see that the pulse decreases in amplitude and broadens only slightly in time. In contrast, the decrease in amplitude and broadening are very apparent in the vertical cut of the current density (Fig. \ref{fig:rtjzp}b). Besides, in the first steps of the vertical propagation, there are two pulses that finally become one after about $t=0.25\tau_F$. This secondary pulse that follows the first agrees with the hypothesis that was proposed in the Sec. \ref{sec:sec3.1} to explain the additional bump in Fig. \ref{fig:nulldiag}b. According to our hypothesis, after the diffusive phase, a pair of FMS pulses is launched from the edge of the diffusion region in opposite directions, outwards from and inwards to the null point. When the inward pulses reach the null, they travel through it and follow behind the opposite outwards propagating pulses, until the second pulse catches up with the first and then they eventually merge into one.

The same cuts are shown for the plasma pressure (Fig. \ref{fig:rtjzp}c and d). In the horizontal cut, we see the main positive pulse, followed by a negative blob which accounts for the inward encroachment of the flow. The pulse in the vertical cut shows the same two-pulse merging observed in the current density.

To finish our analysis, the last question is, does the pulse as a whole carry a net current with it? In order to evaluate this, we define the quantity $C(r,t)$ as the integrated perturbed current inside a circle of radius $r$ at a time $t$,
\begin{equation}
C(r,t)=\int_0^{2\pi}\int_0^r\!\Delta j_z(r,\theta,t)\,r{\rm d}r{\rm d}\theta\;, \label{circle}
\end{equation}
where $\Delta j_z$ represents the perturbation in current density, i.e. after subtracting the initial state. This quantity is equivalent to $C(r,t)$ in LP07 except that here we integrate over the azimuthal direction instead of assuming one-dimensionality of the problem, and we consider the perturbation, $\Delta j_z$ ,as opposed to the absolute value, $j_z$.

This function depends on one spatial coordinate and on time. Fig. \ref{fig:circle} shows plots of $C(r)$ for the same seven times as considered in Fig. \ref{fig:rtjzp}. For a given time, the function $C(r)$ represents the total current contained in circles of increasing radius $r$.

The most central regions are extremely negative. This is because we are dealing here with differences in current density, and this region corresponds to the initial high current density peak. Then, there is a layer of positive current that broadens. This corresponds to the 8-petal flower in Fig. \ref{fig:2Dmaps}a, where the positive (red) petals dominate on the inside while the negative (blue) petals dominate outside. However, $C(r)$ vanishes after this layer, and hence, the whole structure has no net current. It can be seen from the plots that both before and after crossing the propagating pulse, the value of $C(r)$ is identically zero. This shows that the propagating wave carries no net current. The consequences of this result are very important, if we take into account the following:

In a closed system where the boundaries are in an unperturbed equilibrium, the integrated current within the system must remain constant, because
\begin{equation}
\frac{\rm d}{{\rm d}t}\oint_C\!{\bf B}\cdot{\bf dl} = \frac{\rm d}{{\rm d}t}\int_S\!\boldnabla\times{\bf B}\cdot{\bf ds} = \frac{\rm d}{{\rm d}t}\int_S\!{\bf j}\cdot{\bf ds}={\bf 0}\;. \label{net}
\end{equation}
Here, $C$ represents the closed boundary of the system, whose length element is ${\bf dl}$, and $S$ represents the domain enclosed by $C$, whose surface element is ${\bf ds}$. The first part of Eq. (\ref{net}) is zero because the boundaries of the system are in equilibrium. Therefore, at least while the propagating wave has not reached the boundaries, the integrated current within our domain must remain constant (this has been verified numerically).

Hence, the fact that the travelling pulse does not carry any net current with it, implies that the decrease in magnetic energy occurs purely through the redistribution of the central current in the region about the null, rather than through a decrease in the total current.


\section{Discussion} \label{sec:sec4}

Many analytical models for potential and force-free two-dimensional current sheets \cite{Bungey95} are available as starting points for reconnection studies in the vicinity of a magnetic X-point. Only in the last few years have a few authors attempted to give a description of the non-force-free solution to the relaxation of magnetic environments under the effect of a plasma pressure \cite{Rastatter94,Craig05,Pontin05,Fuentes11}, but none of these cases have studied the subsequent current dissipation due to the sudden onset of magnetic reconnection in a finite diffusion region, i.e. what happens when the reconnection is not driven but occurs spontaneously. The closest study was by LP07. They provide an analytical, simplified model of the consequences of magnetic reconnection at a potential Green's current sheet in which pressure effects are neglected.

The approach taken in this paper has been to start from a genuine MHS numerical equilibrium, focusing on the region very near the null where the magnetic field is small, and hence the effects of plasma pressure are important. Here, we have studied numerically the consequences of spontaneous magnetic reconnection in a non-force-free equilibrium current layer (see FF11 for details), after a sudden onset of the resistivity, localised in a region about the null. The initial current density is greatly enhanced near the null, but also extends along the four separatrices. The form of the resistivity is such that it is only non-zero above a certain value of the current density, namely $j_{crit}$. Hence, diffusion will occur only until the current density drops below that level. Due to the fact that the initial diffusion region (where $j_z>j_{crit}$) is very small, the perturbation in the field and the plasma caused by the reconnection is very small, compared to the initial distribution of both the Lorentz forces and plasma pressure gradients.

The evolution of the system is divided into a diffusive phase and a propagating wave phase. During the first phase, the maximum current (at the null) is rapidly decreased below the level of $j_{crit}$. Within this short time, some magnetic energy is converted into both internal and kinetic energy. The overall energetics of our numerical experiments are very different from LP07. In their experiment, most of their magnetic energy is directly converted into kinetic energy in the form of a propagating FMS wave, since their model has vanishing plasma beta. In the end the net current in the initial current sheet is almost entirely redistributed by this process, and the free magnetic energy drops by a significant fraction. According to their model, later on in the evolution of the field, this FMS wave would damp its energy far from the diffusion region.

A new scenario is presented in which pairs of oppositely directed planar pulses are launched from the top and bottom edges of the expanded diffusion region, towards the null and outwards from the null. The pulses travelling towards the null quickly pass through it and follow their opposite outward partner until merging with it. At the same time, two semicircular pulses are launched from the left and right vertices of the diffusion region.

The reconnection velocity flow pattern, inwards above and below the null and outwards to the left and right, defines the structure of the plasma pressure and current density propagating pulses. On the one hand, the plasma pressure pulse propagates a pressure deficit in the regions above and below the null (where there is an inward velocity flow) and a pressure enhancement to the right and left of the null (where there is an outward velocity flow). On the other hand, the magnetic perturbation shows mainly a $m=4$ mode, whose Lorentz forces match with the required reconnection flow pattern. This differs from the LP07 model, in which the $m=0$ mode is assumed to dominate. This difference arises naturally for two main reasons. Firstly, in the LP07 calculations, the initial current is concentrated at a single point, and in order to make progress analytically, they assume radial symmetry from the outset (even though the system is not actually radial). In our numerical experiment, our current sheet is elongated (although still short) and the obvious non-radial symmetry of the X-type magnetic field comes into play. Secondly, in our numerical experiments, we find four different pulses emanating from the diffusion region and then expanding throughout the non-homogeneous medium. This suggests that a dominating $m=0$ radial mode for the magnetic field perturbation could never exist, and rather we obtain a $m=4$ mode, which arises naturally from the geometry of the magnetic field and from the reconnection flow pattern.

Also, the current perturbation carries no net current with it, which differs from LP07 due to our different assumptions about the magnetic diffusivity (namely, that it vanishes below a critical current density). This result, added to the fact that, while the perturbation does no reach the boundaries, the integrated current within the domain must remain constant, as stated in Eq. (\ref{net}), indicates that the decrease in magnetic energy occurs purely through a redistribution of the central current in the region about the null, rather than through a decrease in the total current.

An interesting feature of LP07 is that after the FMS wave has escaped the central current sheet, a persistent X-point electric field continues flux transfer, with little energy dissipation. This feature is not present in our experiments, since we switch reconnection off when the current density is small. For comparison, an experiment with uniform resistivity was also run. The results from such an experiment are different from both the main experiments in this paper and the analytical work of LP07, in two ways. Firtly, there is no clear and (near to) concentric wave expansion, because of the additional diffusion of the enhanced current along the four separatrices and the background current; Secondly, the system eventually reaches the potential configuration, i.e. a simple hyperbolic X-point.

In the numerical experiments presented in this paper, the amplitude of the initial current layer is small, and the diffusion region is short and thin. Hence, the amount of available magnetic energy is tiny. This fact makes the propagating pulses almost negligible. In a more realistic situation, we would expect to see viscous damping of the perturbations, causing a further heating of the plasma away from the diffusion region. This cannot be seen in our experiments because, both the amplitude of the perturbations and the viscosity are tiny, the viscous damping does not occur before the pulses reach the boundary of our system.

In addition, the initial internal energy is about 4 times larger than the initial magnetic energy, and the plasma beta in our domain is large. Therefore, the characteristic speeds of the slow and fast modes are very similar. This feature has the disadvantage of not allowing us to distinguish between fast and slow modes, but has the advantage of making the propagation of the pulses quasi-circular, facilitating calculations such as the net propagation of current density.

In a further paper, we will investigate the same reconnection process in a low beta scenario where the initial current density at the null is about two orders of magnitude larger, and the initial current layer is thinner and longer. In these further experiments, the fast and slow magnetosonic speeds will be significantly different so we hope to be able to distinguish between these two wave modes. Also, we will look for energy conversion far from the diffusion region, due to viscous damping of the waves, and we will evaluate if the portioning of energies depends on the initial choice of plasma parameters.

The consequences of the present study are of potential importance for the chromospheric and coronal heating problem. Under the assumptions taken in this paper, the spontaneous reconnection of a current layer through the relaxation of magnetic fields driven, for example, by slow footpoint motions, may act as a source for direct plasma heating in non-zero beta environments. Similar experiments for three-dimensional magnetic null point reconnection are to be carried out as a natural continuation of the present study.


\section*{Acknowledgements}

The authors would like to thank Dr. I. De Moortel for useful discussions. Computations were carried out on the UKMHD consortium cluster funded by STFC and SRIF.


\bibliographystyle{revtex4}

\begin{thebibliography}{60}

  \bibitem[{{Walsh and Ireland}(2003)}]{Walsh03}
    R.~W. Walsh and J. Ireland, {A\&A Rev.} {\bf 12}, 1-41 (2003).
  \bibitem[{{Klimchuk}(2006)}]{Klimchuk06}
    J.~A. Klimchuk, {Sol. Phys.} {\bf 234}, 41-77 (2006).
  \bibitem[{{Hood}(2010)}]{Hood10}
    A.~W. Hood, {in Lecture Notes in Physics, Berlin Springer Verlag} {\bf 793}, 109 (2010).
  \bibitem[{{Ugarte-Urra et~al.}(2007)}]{Ugarte07}
    I. Ugarte, H.~P. Warren and A.~R. Winebarger, {\apj} {\bf 662}, 1293-1301 (2007).
  \bibitem[{{Longcope and Parnell}(2009)}]{Longcope09}
    D.~W. Longcope and C.~E. Parnell, {Sol. Phys.} {\bf 254}, 51-75 (2007).
  \bibitem[{{Masson et~al.}(2009)}]{Masson09}
    S. Masson, E. Pariat, G. Aulanier and C.~J. Schrijver, {\apj} {\bf 700}, 559-578 (2009).
  \bibitem[{{Parker}(1972)}]{Parker72}
    E.~N. Parker, {Astrophys. J.} {\bf 174}, 499 (1972).
  \bibitem[{{van Ballegooijen}(1985)}]{vanBallegooijen85}
    van Ballegooijen, {Astrophys. J.} {\bf 298}, 421-430 (1985).
  \bibitem[{{Galeev and Sagdeev}(1984)}]{Galeev84}
    A.~A. Galeev and R.~Z. Sagdeev, in {\it Handbook of Plasma Physics} (edited by A. A. Galeev and R. N. Sudan, 1984) {\bf 2}, 271-303.
  \bibitem[{{Raadu and Rasmussen}(1988)}]{Raadu88}
    M.~A. Raadu and J.~J. Rasmussen, {Astrophys. Space Sci.} {\bf 144}, 43-71 (1988).
  \bibitem[{{Yamada et~al}(1997)}]{Yamada97}
    M. Yamada, H. Ji, S. Hsu, T. Carter, R.~M. Kulsrud, N. Bretz, F. Jobes, Y. Ono and F.~W. Perkins, {Phys. Plasmas} {\bf 4}, 1937-1944 (1997).
  \bibitem[{{Antiochos and Sturrock}(1982)}]{Antiochos82}
    S.~K. Antiochos and P.~A. Sturrock, {Astrophys. J.} {\bf 254}, 343-348 (1982).
  \bibitem[{{B\'arta et~al.}(2011)}]{Barta11}
    M. B\'arta, J. B\"uchner, M. Karlick\'y and J. Sk\'ala, {Astrophys. J.} {\bf 737}, 24 (2011).
  \bibitem[{{Galsgaard and Nordlund}(1996)}]{Galsgaard96}
    K. Galsgaard and \AA. Nordlund, {J. Geophys. Res.} {\bf 101}, 13445-13460 (1996).
  \bibitem[{{Longcope and Strauss}(1994)}]{Longcope94}
    D.~W. Longcope and H.~R. Strauss, {Astrophys. J.} {\bf 426}, 742-757 (1994).
  \bibitem[{{Craig and Henton}(1994)}]{Craig94}
    I.~J.~D. Craig and S.~M. Henton, {Astrophys. J.} {\bf 434}, 192-199 (1994).
  \bibitem[{{Petschek}(1964)}]{Petschek64}
    H.~E. Petschek, {NASA Special Publication} {\bf 50}, 425 (1964).
  \bibitem[{{Biskamp}(1986)}]{Biskamp86}
    D. Biskamp, in {\it Magnetic Reconnection and Turbulence} (edited by M. A. Dubois, D. Gr\'esellon and M. N. Bussac, 1964), 19.
  \bibitem[{{Priest and Forbes}(1986)}]{Priest86}
    E.~R. Priest and T.~G. Forbes, {J. Geophys. Res.} {\bf 91}, 5579-5588 (1986).
  \bibitem[{{Longcope and Priest}(2007)}]{Longcope07}
    D.~W. Longcope and E.~R. Priest, {Phys. of Plasmas} {\bf 14}, 122905 (2007).
  \bibitem[{{Green}(1965)}]{Green65}
    R.~M. Green, in {\it Stellar and Solar Magnetic Fields} (edited by R. Lust, 1965) {\bf 22}, 389.
  \bibitem[{{Forbes et~al.}(1982)}]{Forbes82}
    T.~G. Forbes, E.~R. Priest and A.~W. Hood, {Journal of Plasma Physics} {\bf 27}, 157-176 (1982).
  \bibitem[{{Fuentes-Fern\'andez et~al.}(2011)}]{Fuentes11}
    J. Fuentes-Fern\'andez, C.~E. Parnell and A.~W. Hood, {A\&A} {\bf 536}, A32 (2011).
  \bibitem[{{Fuentes-Fern\'andez et~al.}(2010)}]{Fuentes10}
    J. Fuentes-Fern\'andez, C.~E. Parnell and A.~W. Hood, {A\&A} {\bf 514}, A90 (2010).
  \bibitem[{{Arber et~al.}(2001)}]{Arber01}
    T.~D. Arber, A.~W. Longbottom, C.~L. Gerrard and, A.~M. Milne, {Computational Phys.} {\bf 171}, 151-181 (2001).
  \bibitem[{{Fontenla}(1993)}]{Fontenla93}
    J. M. Fontenla, {Astrophys J.} {\bf 419}, 837-854 (1993).
  \bibitem[{{Rast\"atter et~al.}(1994)}]{Rastatter94}
    L. Rast\"atter, A. Voge and K. Schindler, {Phys. Plasmas} {\bf 1}, 3414 (1994).
  \bibitem[{{Craig and Litvinenko}(2005)}]{Craig05}
    I.~J.~D. Craig and Y.~E. Litvinenko, {Phys. Plasmas} {\bf 12}, 032301 (2005).
  \bibitem[{{Pontin and Craig}(2005)}]{Pontin05}
    D.~I. Pontin and I.~J.~D. Craig, {Phys. Plasmas} {\bf 12}, 072112 (2005).
  \bibitem[{{Bungey and Priest}(1995)}]{Bungey95}
    T.~N. Bungey and E.~R. Priest, {A\&A} {\bf 293}, 215-224 (1995).
  \bibitem[{{Yokoyama and Shibata}(1994)}]{Yokoyama94}
    T. Yokoyama and K. Shibata, {Astrophys. J. lett.} {\bf 436}, L197-L200 (1994)
  \bibitem[{{Ugai et~al}(2003)}]{Ugai03}
    M. Ugai, K. Kondoh and T. Shimizu, {Phys. of Plasmas} {\bf 10}, 357 (2003).
  \bibitem[{{Ugai and Zheng}(2005)}]{Ugai05}
    M. Ugai and L. Zheng, {Phys. of Plasmas} {\bf 12}, 092312 (2005).
  \bibitem[{{Uzdensky}(2005)}]{Uzdensky05}
    D. A. Uzdensky, {Astrophys. J.} {\bf 587}, 405 (2005).
  \bibitem[{{Roussev et~al}(2002)}]{Roussev02}
    I. Roussev, K. Galsgaard and P.~G. Judge, {A\&A} {\bf 382}, 639-649 (2002).
  \bibitem[{{Krasnoselskikh et~al}(2002)}]{Krasnoselskikh02}
    V. Krasnoselskikh, O. Podladchikova, B. Lefebvre and N. Vilmer, {A\&A} {\bf 382}, 699-712 (2002).

\end{thebibliography}


\end{document}